\documentclass[aps,pra,reprint,superscriptaddress,footinbib,]{revtex4-2}

\usepackage[T1]{fontenc}			
\usepackage[utf8]{inputenc}			
\usepackage[english]{babel}			
\usepackage[protrusion=false]{microtype}	

\usepackage{xcolor} 						
\definecolor{red}{rgb}{1,0,0}				
\definecolor{blue}{rgb}{0,0,1}				
\definecolor{black}{rgb}{0,0,0}				

\usepackage{soul} 
\definecolor{hlyellow}{rgb}{0.95,0.95,0}
\definecolor{hlgreen}{rgb}{0,0.95,0}
\sethlcolor{hlyellow}
\soulregister \ref 7
\soulregister \cite 7
\soulregister \citet 7
\soulregister \footnote 8
\soulregister {\ } 7
\soulregister \figref 7
\soulregister \tabref 7
\soulregister \secref 7
\soulregister \supref 7
\soulregister \refref 7
\soulregister \eqref 7
\soulregister \supmat 7

\usepackage{amsmath} 
\usepackage{amssymb,amsfonts,mathrsfs} 
\usepackage{array, dcolumn} 

\usepackage{graphicx} 
\usepackage{hyperref} 
\definecolor{dullmagenta}{rgb}{0.4,0,0.4} 
\definecolor{darkblue}{rgb}{0,0,0.4}
\definecolor{medblue}{rgb}{0,0,0.6}
\definecolor{lightblue}{rgb}{0,0,0.8}
\hypersetup{
	colorlinks=true, 		
	breaklinks=true,		
	linkcolor=lightblue,
	citecolor=lightblue,
	urlcolor=lightblue,
	filecolor=dullmagenta
}
\usepackage[all]{hypcap} 
\usepackage{physics} 
\usepackage{bbold} 

\newcommand{\figref}[1]{Fig.~\ref{#1}} 
\newcommand{\tabref}[1]{Tab.~\ref{#1}} 
\newcommand{\secref}[1]{Sec.~\ref{#1}} 
\newcommand{\refref}[1]{Ref.~\cite{#1}} 
\newcommand{\eqnref}[1]{Eq.~\eqref{#1}} 

\usepackage{upgreek}

\makeatletter 
\newsavebox{\@brx}
\newcommand{\llangle}[1][]{\savebox{\@brx}{\(\m@th{#1\langle}\)}%
  \mathopen{\copy\@brx\kern-0.5\wd\@brx\usebox{\@brx}}}
\newcommand{\rrangle}[1][]{\savebox{\@brx}{\(\m@th{#1\rangle}\)}%
  \mathclose{\copy\@brx\kern-0.5\wd\@brx\usebox{\@brx}}}
\makeatother

\newcommand{\mr}{\mathrm}
\newcommand{\sbra}[1]{\llangle #1|}
\newcommand{\sket}[1]{| #1 \rrangle}

\newcommand{\supmat}{Supplementary Material}
\newcommand{\supref}[1]{App.~\ref{#1}} 

\begin{document}

\title{Resonant two-qubit gates for fermionic simulations with spin qubits}

\newcommand{\zrl}{IBM Research Europe -- Zurich, Säumerstrasse 4, 8803 Rüschlikon, Switzerland}
\newcommand{\yorktown}{IBM Quantum, T.\,J.\,Watson Research Center, Yorktown Heights, NY, USA}

\author{Konstantinos \surname{Tsoukalas}}
\thanks{These authors contributed equally to this work.}
\affiliation{\zrl}
\author{Alexei \surname{Orekhov}}
\thanks{These authors contributed equally to this work.}
\author{Bence \surname{Hetényi}}
\affiliation{\zrl}
\author{Uwe~von \surname{Lüpke}}
\affiliation{\zrl}
\author{Jeth \surname{Arunseangroj}}
\affiliation{\zrl}
\author{Inga \surname{Seidler}}
\affiliation{\zrl}
\author{Lisa \surname{Sommer}}
\affiliation{\zrl}
\author{Eoin~G. \surname{Kelly}}
\affiliation{\zrl}
\author{Leonardo \surname{Massai}}
\affiliation{\zrl}
\author{Michele \surname{Aldeghi}}
\affiliation{\zrl}
\author{Marta \surname{Pita-Vidal}}
\affiliation{\zrl}
\author{Stephen~W. \surname{Bedell}}
\affiliation{\yorktown}
\author{Stephan \surname{Paredes}}
\affiliation{\zrl}
\author{Felix~J. \surname{Schupp}}
\affiliation{\zrl}
\author{Matthias \surname{Mergenthaler}}
\affiliation{\zrl}
\author{Gian \surname{Salis}}
\affiliation{\zrl}
\author{Andreas \surname{Fuhrer}}
\affiliation{\zrl}
\author{Patrick \surname{Harvey-Collard}}
\email[Corresponding author: ]{phc@zurich.ibm.com}
\affiliation{\zrl}

\date{July 17, 2025}

\begin{abstract}
In gate-defined semiconductor spin qubits, the highly tunable Heisenberg exchange interaction is leveraged to implement fermionic two-qubit gates such as CZ and SWAP. However, the broader family of fermionic simulation (fSim) gates remains unexplored, and has the potential to enhance the performance of near-term quantum simulation algorithms. Here, we demonstrate a method to implement the fSim gate set in spin qubits using a single pulse combining baseband and resonant exchange drives. This approach minimizes gate duration and drive amplitude, mitigating decoherence and crosstalk. We validate its effectiveness by realizing a resonant iSWAP gate between two hole spins in germanium, achieving a fidelity of 93.8(5)\% extracted with interleaved randomized benchmarking. Quantum process tomography confirms accurate gate calibration and identifies qubit decoherence as the dominant error source. Our results establish a practical route toward a versatile and efficient two-qubit gate set for spin-based quantum processors. 
\end{abstract}

\maketitle

\section{Introduction}
\label{sec:level1}

A wide class of physical models, such as the Hubbard model and variational ansätze in quantum chemistry, can be described by fermionic Hamiltonians \cite{Kivlichan2018, cade2020fermihubb, Verstraete2009}. Having direct access to the class of fermionic simulation gates $\mathrm{fSim}(\gamma, \zeta) = \mathrm{iSWAP}^{-2\gamma/\pi}\mathrm{CZ}^{\zeta/\pi}$ facilitates fermionic time evolution with linear circuit depth, even in a one-dimensional chain of qubits \cite{Kivlichan2018, Verstraete2009, jiang2024concurrent}. This is because fSim gates preserve the particle-exchange symmetry of the simulated wavefunction under the Jordan–Wigner fermion-to-qubit mapping. These advantages have led to the widespread adoption of the fSim gate in the field of superconducting qubits for both quantum simulation and quantum computing applications \cite{Google2020, Abrams2020,Sung2021,  ganzhorn2019gate, google2020hartree, yun2024one,   arute2020observationseparateddynamicscharge, FluxoniumfSim, Moskalenko2022}.

In spin qubits, (fermionic) two-qubit gates are implemented by leveraging the highly tunable exchange interaction ($J$). The conditional phase gate $\mathrm{CZ}^{\zeta/\pi}$ is realized by adiabatically activating the exchange, most commonly in the $J \ll \Delta E_Z$ regime \cite{watson2018programmable,hendrickx2021four}, where $\Delta E_Z$ is the Zeeman energy difference between the spins. The $\mathrm{CZ}$ gate is the most established two-qubit gate in the field, having achieved fidelities exceeding 99\% on multiple platforms \cite{Xue2022,wang2024operating,steinacker2024300,mkadzik2022precision,thorvaldson2025grover}. On the other hand, by diabatically pulsing $J$ into the $J \gtrsim \Delta E_Z$ regime, swap oscillations can be induced \cite{ni2025swap,petit2022design}. However, the high sensitivity of the spin states to charge noise in this regime, the large circuit bandwidth required and the need for composite rotations has so far made this gate less favorable \cite{petit2022design, ni2025swap, RachponSWAP2014}. Alternatively, swap oscillations can be generated in the low $J$ regime ($\Delta E_Z \gg J$) \cite{Nguyen2023, Philips2023}, by driving $J$ at a frequency resonant with $\Delta E_Z / h$. This technique has demonstrated a coherent $\mathrm{SWAP} = \mathrm{fSim}(-\pi/2, \pi)$ gate with a fidelity of approximately 84\%, but it requires large drive amplitudes (to activate $J$) and an additional $\mathrm{CZ}^{\zeta/\pi}$ gate to calibrate the phase \cite{Sigillito2019}.

For implementing the fSim gate in spin qubits \cite{ni2025diverse}, a particularly promising approach is to combine resonant and baseband exchange pulses \cite{van2022phase}, which together enable simultaneous control over both swap-like exchange dynamics and conditional phase accumulation. This method offers several advantages. First, it avoids the need for large $J$, which would expose the spin states to significant charge noise. Second, the simultaneous application of resonant and baseband pulses can reduce both the gate duration and the drive amplitude, thereby mitigating decoherence and crosstalk during gate operation \cite{undseth2023nonlinear, kelly2023capacitive}. Finally, the additional degree of freedom provided by the drive phase makes the resonant fSim gate compatible with virtual Z rotations \cite{ChenPRR}, unlike its baseband counterparts \cite{ arute2020observationseparateddynamicscharge,ni2025diverse}. Despite these advantages, a comprehensive theoretical and experimental investigation of the parameter space required to implement a general fSim gate using the resonant exchange interaction is still lacking.

In this work, we introduce methods to measure and control the underlying parameters of the resonant $\mathrm{fSim}(\gamma, \zeta)$ gate in spin qubits, implemented using simultaneous resonant and baseband exchange pulses. Single-qubit corrections and the conditional phase are tuned using a combination of Ramsey sequences and partial tomographies. We then apply these techniques to realize a coherent $\mathrm{iSWAP} = \mathrm{fSim}(-\pi/2, 2\pi)$ gate between two hole spins in germanium. Gate characterization via interleaved randomized benchmarking yields fidelities of up to 93.8(5)\%. To investigate the dominant error sources, we perform quantum process tomography, which indicates that the gate performance is limited by decoherence while confirming the accuracy of the calibration. 

\begin{figure}[tbp]
\centering
\includegraphics[scale=1]{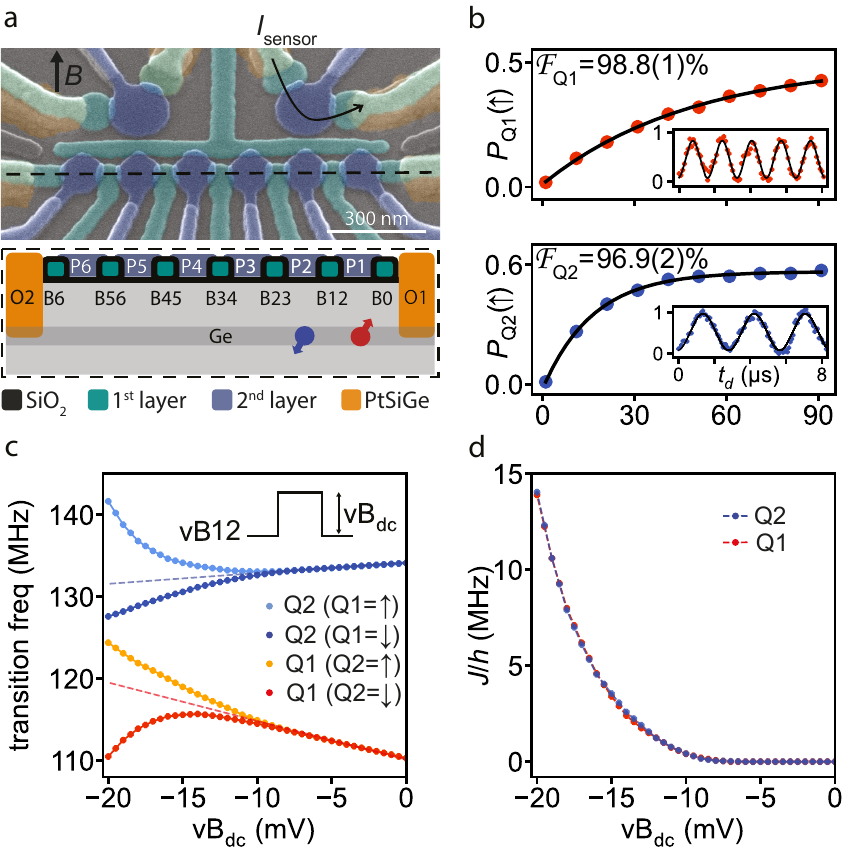}
\caption{
\textbf{Single qubit gates and exchange calibration.} 
\textbf{(a)}~A false-colored scanning electron micrograph of a six-QD device nominally identical to the one used in the experiments. Below, a schematic of the device cross section along the dashed black line with the blue and red spins indicating the position of the two hole spin qubits. 
\textbf{(b)}~Randomized benchmarking of the single-qubit Clifford gates for each qubit, along with their Rabi oscillations (inset). Average physical gate fidelities ($X_{\pi},X_{\frac{\pi}{2}}$) of $\mathcal{F}_\mathrm{Q1} = 98.8(1)\%$ and $\mathcal{F}_\mathrm{Q2} = 96.9(2)\%$ were extracted for Q1 and Q2, respectively. 
\textbf{(c)}~Measurement of the frequency of each qubit conditional on the state of the other while applying a square pulse of amplitude $\mr{vB}_\mathrm{\text{dc}}$ on the barrier. By linearly extrapolating the average of the Larmor frequency of each qubit we extract a barrier pulse-induced frequency shift of $\frac{\partial f_{\mathrm{Q1}}}{\partial \mr{vB}_\mr{\text{dc}}} = -0.44$
\,MHz/mV and $\frac{\partial f_{\mathrm{Q2}}}{\partial \mr{vB}_\mr{\text{dc}}} =0.13$\,MHz/mV for Q1 and Q2, respectively. 
\textbf{(d)}~The exchange interaction extracted from the Larmor frequency difference of Q1 (red) and Q2 (blue) conditional on the state of the other spin. }
\label{fig1}
\end{figure}

\section{Device properties}

We employ a six-quantum-dot (QD) device, identical to Ref.\,\cite{tsoukalas2025dressed} (\figref{fig1}(a)), fabricated using a two-gate-layer process on a Ge/SiGe heterostructure. The first layer, barrier (B) gates, controls the coupling between QDs, and the second layer, plunger (P) gates, tunes the chemical potential of each QD, while an annealed Pt layer defines the hole reservoirs \cite{hendrickx2024sweet,tsoukalas2024prospects,tsoukalas2025dressed}. We focus on the double QD (DQD) formed under P1 and P2. The charge occupation of the DQD is measured by monitoring the current of a capacitively coupled nearby sensing QD ($I_\mr{sensor}$). By adiabatically pulsing from the S(2,0) to the middle of the (1,1) charge regions, while also increasing vB12, we initialize two spins in the $|{\downarrow\downarrow}\rangle$ ground state with $J \approx 0$. To determine the spin state of each qubit we perform the reverse adiabatic pulse followed by latched Pauli spin blockade (PSB) readout \cite{harvey2018high, kelly2025identifying}, which distinguishes between the $|{\downarrow\downarrow}\rangle$ state and all other states. By flipping the spins of Q1 (spin under P1) or Q2 (spin under P2), any two-spin state $\ket{\mathrm{Q1,Q2}}$ can be projected to $\ket{\downarrow\downarrow}$ for readout. We perform time-resolved averaged current readout unless single-shot readout is indicated (\supref{app:setup}). 

The device operates in an in-plane magnetic field of $B=20$\,mT, resulting in Larmor frequencies $f_\mr{Q1}=110.3$\,MHz and $f_\mr{Q2}=134.1$\,MHz for the two qubits. Each spin is driven by modulating its g-tensor through an oscillating electrical signal applied to the respective plunger gate \cite{hendrickx2024sweet}. The drive amplitude of each qubit is set to 6\,mV, resulting in spin flip times (\figref{fig1}(b) insets) of 900\,ns and 1400\,ns for Q1 and Q2, respectively. By applying a Ramsey sequence (see \supref{app:Coherence times}) we extract coherence times of $T^*_\mr{2,Q1}=2.6\,$\textmu s and $T^*_\mr{2,Q2}=1.8\,$\textmu s. To characterize the single qubit gate fidelity, we use randomized benchmarking, which yields fidelities of $\mathcal{F}_\mr{Q1} = 98.8(1)\%$ and $\mathcal{F}_\mr{Q2} =96.9(2)$\% as shown in \figref{fig1}(b).

High-fidelity two-qubit gates require precise tuning of the exchange interaction between neighboring spins. We control $J$ with the voltage applied to the virtualized barrier vB12, allowing us to maintain operation at the symmetric exchange point \cite{reed2016reduced}. The magnitude of $J$ can be extracted from the difference between the frequencies of each qubit conditional on the state of the other qubit. 
The frequencies of the two hole spins are probed by performing rf spectroscopy while applying a dc voltage pulse to vB12 of amplitude $\mr{vB}_\mr{\text{dc}}$. 
The measured transition frequencies for both qubits are presented in \figref{fig1}(c). We observe that until $\mr{vB}_\mr{\text{dc}}\approx-8$\,mV, the excitation frequency of each spin remains largely independent of the state of the other. Above this value, each frequency splits into two branches, with a difference between them equal to $J$. In addition, $\mr{vB}_\mr{\text{dc}}$ also affects the shape and location of the QDs, inducing shifts to their g-factors which are independent of the state of the other spin and depend approximately linearly on $\mr{vB}_\mr{\text{dc}}$.   
In \figref{fig1}(d) we plot the extracted $J$, recognizing the characteristic exponential dependence of the exchange on $\mr{vB}_\mr{\text{dc}}$.

\begin{figure}[tbp]
\includegraphics[width=\columnwidth]{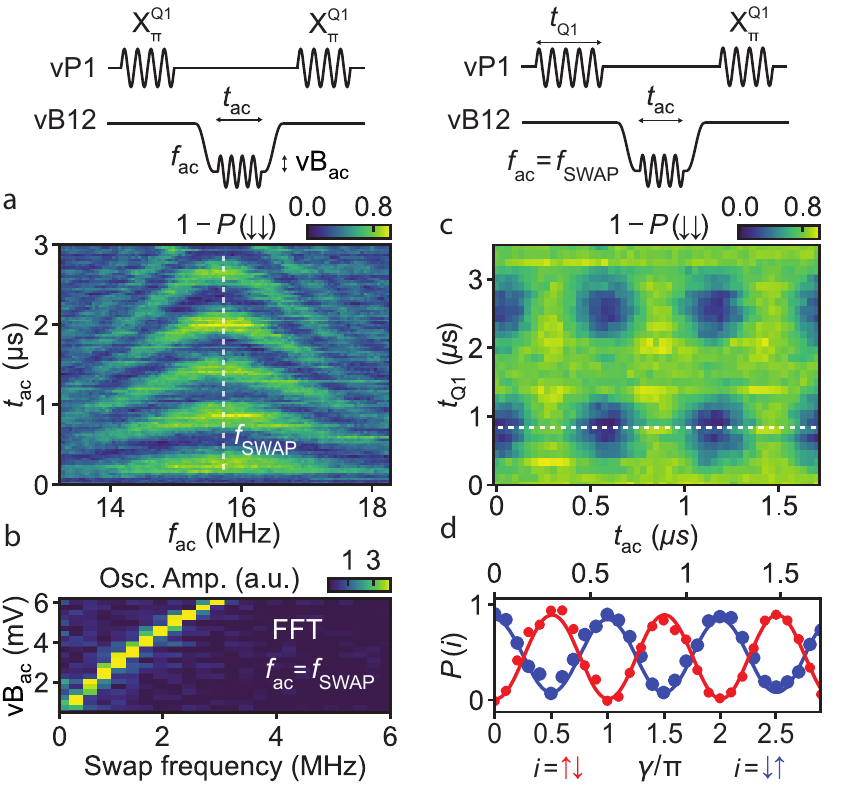}
\caption{
\textbf{Resonant Exchange Oscillations.} 
\textbf{(a)}~Swap oscillations induced by superposing an oscillating signal, of frequency $f_\mr{\text{ac}}$ and of amplitude $\mr{vB}_\mr{\text{ac}}=4$~mV, with a baseband barrier pulse of amplitude $\mr{vB}_\mr{\text{dc}}$ with ramp time $t_\mr{ramp} = 100$\,ns. 
\textbf{(b)}~Fast Fourier transform (FFT) reveals the speed of the swap, driven at $f_\mr{ac}=f_\mr{SWAP}$, for different drive amplitudes $\mr{vB}_\mr{ac}$. 
\textbf{(c)}~Rabi drive of Q1 followed by a resonant exchange drive. 
\textbf{(d)}~Population transfer between $|{\uparrow\downarrow}\rangle$ and $|{\downarrow\uparrow}\rangle$ through resonant swap oscillations along the white dashed line in (c). In the lower x-axis the corresponding value of the $\mr{iSWAP}^{-2\gamma/\pi}$ component of the $\mr{fSim}(\gamma, \zeta)$ gate is indicated. }
\label{fig2}
\end{figure}

\section{Exchange Oscillations}

Resonant driving of the exchange interaction offers a mechanism to achieve a swap-like operation between two adjacent spins \cite{Sigillito2019,van2022phase}. In this scheme, exchange is modulated at frequency $f_\mr{SWAP} = \sqrt{\Delta E_z^2 + J^2}/h$ \cite{takeda2020resonantly,saez2025exchange}. We demonstrate resonant swap oscillations by preparing $|{\uparrow\downarrow}\rangle$ and applying a baseband (dc) pulse of amplitude $\mr{vB}_\mr{\text{dc}}=-14$\,mV and ramp time $t_\mr{ramp} = 100$\,ns. The superimposed oscillating (ac) signal has amplitude $\mr{vB}_\mr{\text{ac}}=4$\,mV and frequency $f_\mr{\text{ac}}$ and is also applied on vB12. When $f_\mr{\text{ac}} \simeq f_\mr{SWAP}$, the characteristic Rabi chevron pattern observed is shown in \figref{fig2}(a).

To investigate the dependence of the swap oscillation speed on the drive amplitude, we set $f_\mr{ac} = f_\mr{SWAP}$ and vary $\mr{vB}_\mr{ac}$. The Fourier transform of the measured data, shown in \figref{fig2}(b), reveals a nonlinear dependence, which originates from the nonlinear relationship between $J$ and the barrier voltage, as depicted in \figref{fig1}(d). Throughout this study, we used $\mr{vB}_\mr{ac}$ values of 4 and 5.3\,mV (see \supref{app:proc_tomo}).

In \figref{fig2}(c), we demonstrate population transfer between the spin states. First, a Rabi drive applied to Q1 for a duration $t_\mathrm{Q1}$, followed by a resonant exchange drive on vB12 for a duration $t_\mathrm{\text{ac}}$ (\figref{fig2}(c)). Along the dashed white line in \figref{fig2}(c), complementary oscillations of the $|{\uparrow\downarrow}\rangle$ and $|{\downarrow\uparrow}\rangle$ states are observed, measured using $\mr{X}^\mr{Q1}_{\pi}$ and $\mr{X}^\mr{Q2}_{\pi}$ readout pulses, respectively (\figref{fig2}(d)). This indicates the continuous transfer of excitation (swap oscillations) between Q1 and Q2. Within the framework of the $\mr{fSim}(\gamma, \zeta)$ gate, this corresponds to a variation of the $\gamma$ parameter, as indicated on the lower x-axis axis of \figref{fig2}(d).

For the implementation of the fSim gates, we explore two pulsing schemes that differ in the ramp time of the dc barrier pulse, with $t_\mathrm{ramp} = 5$\,ns and $t_\mathrm{ramp} = 100$\,ns. While the shorter ramp time leads to a faster gate operation, reducing exposure to decoherence, it is subject to weak secondary oscillations which originate from the diabatic change of basis states, caused by the rapid activation of $J$. In contrast, the longer ramp time effectively suppresses these oscillations, indicating a transition into the adiabatic regime. Consequently, we term the $t_\mathrm{ramp} = 5$\,ns case as diabatic and the $t_\mathrm{ramp} = 100$\,ns case as adiabatic. A detailed discussion of non-adiabatic effects and the rationale behind our gate parameter choices is provided in \supref{app:basebandSWAP}.

\section{fSim gate calibration}

\begin{figure*}[tbp]
\centering
\includegraphics[width=17.5cm]{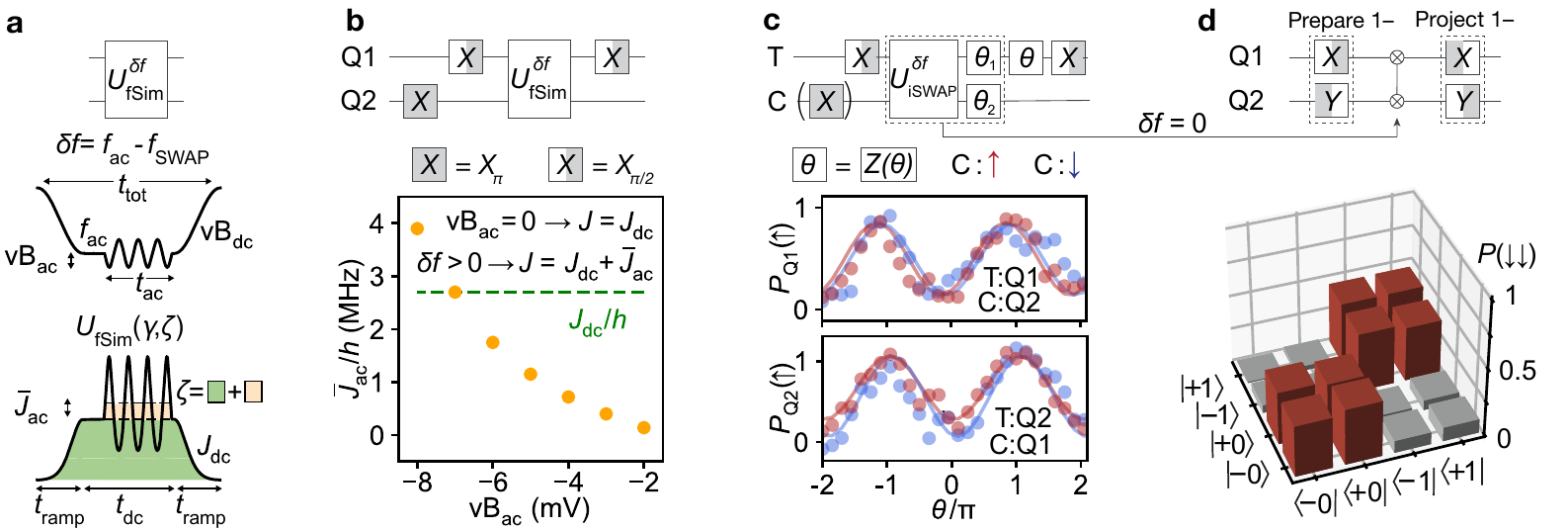}
\caption{
\textbf{Resonant iSWAP phase calibration.} 
\textbf{(a)}~Schematic of the pulse shape applied to vB12 (upper trace) to realize a resonant fSim gate. An ac drive of frequency $f_\mathrm{\text{ac}}$ is superimposed with a Tukey-shaped baseband pulse. The baseband pulse results in an exchange of ${J}_{\mathrm{\text{dc}}}$ (lower trace), while the induced exchange modulation  results in an additional exchange component $\overline{J}_{\mathrm{\text{ac}}}$ arising from the non-linear dependence on vB12. The conditional phase $\zeta$ of the $\mr{fSim}(\gamma, \zeta)$ gate is generated by the addition of the two conditional phases. 
\textbf{(b)}~Extraction of $\overline{J}_{\mathrm{\text{ac}}}$ for different drive amplitudes by performing a Ramsey sequence while driving off-resonantly to avoid swap rotations. 
\textbf{(c)}~Measurement of the phase of the target qubit for both basis states of the control qubit after applying an adiabatic off-resonant pulse. A total phase of $2\pi$ confirms the iSWAP nature of these drive parameters.  
\textbf{(d)}~An example of a partial process tomography, used to fine-tune the correction phase $\alpha_2$. An equal height of the columns in the diagonal 2x2 blocks indicates a good calibration of the $\alpha_2$ correction (see \supref{app:subsec2}). The tomography data is obtained using single-shot readout (\supref{app:setup}). } 
\label{fig3}
\end{figure*}

The fSim interaction is obtained by applying a baseband pulse of duration $t_{\mathrm{tot}} = t_\mathrm{dc} + 2t_{\mathrm{ramp}}$ and a superimposed ac drive of duration $t_{\mathrm{ac}}$ on vB12 (\figref{fig3}(a)). While the ac component of $J$ induces swap-like oscillations (parametrized by $\gamma$), the integrated value of $J$ results in a conditional phase accumulation $\zeta$. The combination of these two effects results in the unitary $U_{\text{fSim}}$ in the two-qubit Hilbert space (\supref{app:Unitary evolution}). The complete unitary of the general fSim gate is then obtained by: $\mathrm{fSim}(\gamma, \zeta)=Z_{\mathrm{1}}(\alpha_1)Z_{\mathrm{2}}(\alpha_2)U_\mathrm{fSim}(\gamma, \zeta)$
where $Z_1(\alpha_1)$, $Z_2(\alpha_2)$ are single-qubit virtual phase corrections with corresponding phases $\alpha_1$ and $\alpha_2$. We note that a drive phase update is additionally required. 

We now introduce a calibration method for the $\text{fSim}$ gate and use it to tune up $\text{iSWAP} = \text{fSim}(-\pi/2, 2\pi)$. The calibration procedure is structured as follows: First, we tune the duration and amplitude of the drive to obtain the desired population exchange $\gamma$. Then, $t_{\mathrm{dc}}$ is adjusted to reach a target conditional phase $\zeta$. In this step, calibration errors arising form the diabatic activation of $J$ are considered small and thus neglected. Finally, we perform a sequence of state preparations and measurements in different bases (partial tomography), to fine-tune $\zeta$ and the single-qubit phase corrections $\alpha_1, \alpha_2$.

The conditional phase $\zeta$ is given by 
\begin{equation} \label{eq:zeta}
    \zeta  =  \int_{t_{\mathrm{tot}}} \frac{J_\mathrm{dc}(t)}{\hbar} \, \mathrm{d}t  +  \frac{\overline{J}_{\mathrm{\text{ac}}}}{\hbar}\,t_{\mathrm{ac}},
\end{equation}
where $J_{\text{dc}}(t)$ is the exchange arising from the baseband pulse and $\overline{J}_{\mathrm{\text{ac}}}$ is the average exchange generated by the ac drive, a result of the nonlinear dependence of $J$ on the gate voltage (\figref{fig3}(a)). In order to isolate the accumulation of the conditional phase from the swap oscillation, we employ off-resonant driving. We can thus induce an average $J$ close to the one of the resonant gate without driving swap oscillations, allowing for the measurement of the exchange with a Ramsey sequence (\supref{app:Gatetuning}) \cite{Xue2022}. We denote the unitary of the off-resonant drive $U^{\updelta f}_{ \text{fSim}}$, where $\updelta f = f_\mathrm{ac}-f_\mathrm{SWAP}$.
In \figref{fig3}(b) we show the extracted average ac and dc contributions to $J$ for a baseband amplitude of $\mr{vB}_\mathrm{\text{dc}}=-14$\,mV and varying ac drive amplitudes, measured with the Ramsey sequence depicted above the plot. Here, the off-resonant drive frequency is set to $f_\mathrm{ac}=30$\,MHz. 

As seen in \figref{fig1}(c), the qubit frequencies vary with the pulse amplitude $\mr{vB}_{\mathrm{dc}}$, inducing single qubit phases which need to be corrected. We denote the frequency shifts $\Delta_1$ and  $\Delta_2$ for the respective qubits and define the induced phases as  $\theta_1 = 2\pi \int_{t_{\text{tot}}} \Delta_1(t) \, \mathrm{d}t $ and $
\theta_2 = 2\pi \int_{t_{\text{tot}}} \Delta_2(t) \, \mathrm{d}t$. For the general fSim, the single qubit phases applied after the pulse are $\alpha_1= \theta_1 +\zeta/2$ and $\alpha_2=\theta_2+\zeta/2$. We find that the dependence of qubit frequency on gate voltage is sufficiently linear such that we can neglect the change of the average qubit frequency during application of the resonant drive. \figref{fig3}(c) shows a conditional Ramsey experiment with varying phase on the recovery pulse, for an off-resonant driven gate with drive amplitude $5.3$ mV, $\mr{vB}_{\mathrm{dc}}=-12.3$\,mV and $t_{\text{ramp}} = 100$\,ns. This measurement is used to calibrate both  $t_{\text{dc}}$, to bring the two conditional oscillations into phase ($\zeta = 2\pi$), as well as $\theta_1$, $\theta_2$ to compensate for single qubit phases. The off-resonant Ramsey-based approach introduced for calibrating $\zeta$, $\theta_1$ and $\theta_2$ can be used to target an arbitrary fSim parameter regime. 

\begin{figure*}[tbp]
\includegraphics[width=17.6cm]{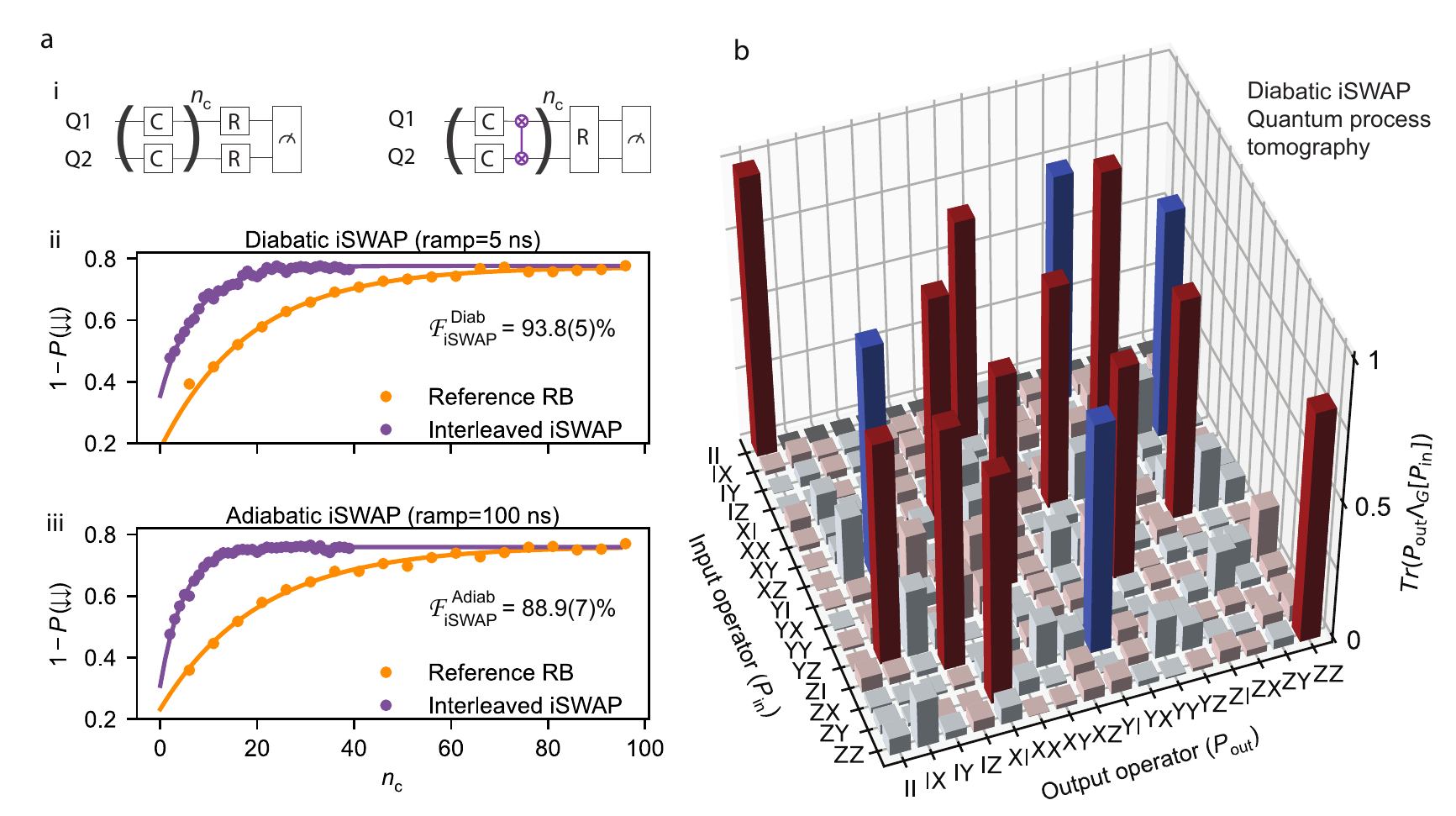}
\caption{\textbf{(a)} i) Circuit diagram for the reference Clifford-based RB sequence (left), and the iSWAP interleaved sequence (right). ii) IRB results of the diabatic iSWAP. iii) Interleaved RB results for the adiabatic iSWAP gate. The probability scales are normalized by fitting shot histograms from neighboring measurements (see \supref{app:setup}). \textbf{(b)} Quantum process tomography of the diabatic iSWAP in PTM representation, where the identity channel has been inverted to eliminate the SPAM errors.}
\label{fig:fig4}
\end{figure*}

Due to the finite detunings $\Delta_1$ and $\Delta_2$, a residual phase $\phi_\mathrm{res}$ accumulates in the anti-parallel subspace during the initial section of the baseband pulse before the start of the drive, affecting the axis of exchange rotation (\supref{app:Unitary evolution}). For a general fSim gate, this can be compensated by adding $\phi_\mathrm{res}$ to the drive phase, realigning the rotation axis along the desired direction. Importantly, once calibrated for a specific fSim gate, this ramp-dependent phase should remain unchanged for an arbitrary fSim. In the specific case of a full (i)SWAP, an alternative approach is to incorporate $\phi_\mathrm{res}$ into the single-qubit corrections, re-defining the phases to be $\alpha_1\rightarrow\alpha_1-\phi_\mathrm{res}$ and $\alpha_2\rightarrow\alpha_2 + \phi_\mathrm{res}$, as shown in \supref{app:Unitary evolution}. 
We adopt this approach for the iSWAP, calibrating $\alpha_1, \alpha_2$ via series of partial tomographies, which may deviate from the estimates in \figref{fig3}(c). An example of such a partial tomography for fine-tuning $\alpha_2$ is shown in \figref{fig3}(d), where the states $\{\ket{-0},\ket{+0},\ket{-1},\ket{+1}\}$  are prepared, followed by the application of a calibrated iSWAP gate, and then a measurement in the basis $\{\ket{0-},\ket{0+},\ket{1-},\ket{1+}\}$. The phase $\alpha_1$ is tuned analogously by preparing $\{\ket{0-},\ket{0+},\ket{1-},\ket{1+}\}$  and measuring $\{\ket{-0},\ket{+0},\ket{-1},\ket{+1}\}$. Further details on the tomography calibrations are provided in \supref{app:subsec2}. The values of all parameters used for the diabatic and adiabatic iSWAP gates can be found in App.\ \tabref{tab:gateParams}.

\section{Gate benchmarking}

The iSWAP gate fidelity is estimated with interleaved randomized benchmarking using the subset of two-qubit Clifford gates comprised of single qubit gates (\figref{fig:fig4}(a)). Whereas the recovery gate for the non-interleaved sequence only includes single-qubit gates, the one for the interleaved case includes iSWAP gates (\supref{app:IRB}). This explains the horizontal offset between the interleaved and non-interleaved curves, and should not influence the decay exponents. The decay curves are fitted with the function $A + B\alpha^{n_{c}}$, where $\alpha$ is the depolarizing parameter and $n_c$ is the number of Cliffords. The fidelity of the interleaved gate is then extracted as $F = 1 - \frac{d-1}{d}\left( 1 - \frac{\alpha_{\mathrm{interleaved}}}{\alpha_{\mathrm{reference}}}\right)$, where $d=4$ is the dimension of the Hilbert space \cite{xue2019benchmarking}. We obtain fidelities of 93.8(5)\% and 88.9(7)\% for the diabatic ($t_\mathrm{tot}=343$\,ns, $t_{\text{ramp}} = 5$\,ns) and adiabatic ($t_\mathrm{tot}=485$\,ns,  $t_{\text{ramp}} = 100$\,ns) gates, respectively. 

To study which processes limit the gate fidelity we perform quantum process tomography \cite{poyatos1997complete,artiles2004invitation}. 
Preparing Pauli eigenstates $\{\ket 0, \ket 1, \ket +, \ket -, \ket{+i}, \ket{-i}\}$ on each of the qubits we perform the iSWAP gate and measure the outcome in the $\{X,Y,Z\}$ bases. Sampling from this set of quantum circuits is sufficient for the tomographic reconstruction of the quantum process. 

Accounting for the SPAM channel extracted from the tomography of the identity gate, we find the SPAM-corrected fidelities for the diabatic and adiabatic iSWAP gates. Although the gate parameters were aimed to be identical to the ones used for the RB, frequent charge jumps required recalibration. Nevertheless, we find higher SPAM-corrected fidelities for the diabatic case ($86.3\%$) compared to the adiabatic one ($81.2\%$). Our method is similar to gate set tomography (GST) in that it separates SPAM errors from gate errors \cite{merkel2013selfconsistent,blume2013robust,Nielsen2021gatesettomography}, but it neglects single qubit gate errors. 
The Pauli transfer matrix (PTM) corresponding to the SPAM-corrected channel is shown in \figref{fig:fig4}(b) for the diabatic gate with details of the analysis in \supref{app:proc_tomo}. The fidelities extracted from the tomographies are limited by {\it (i)} our assumptions on the SPAM channel, and {\it (ii)} the neglected single-qubit gate errors that are significant in our case ($\sim 3\%$). Finally, by removing the decoherence errors we extract the unitary component of the channel and find that calibration errors are on par with single-qubit gate errors (which set the limit of this method), confirming the successful phase calibration of the gate.

\section{Discussion}

We observe that a shorter ramp results in higher gate fidelity, which is consistent with the gate being limited by spin dephasing. This is corroborated by our analysis of a quantum process tomography. Improving gate fidelity could involve both extending the coherence time of single spins and accelerating the gate operation. The coherence time can be extended by an order of magnitude with an optimized magnetic field direction, making use of charge or nuclear noise sweetspots (or sweetlines)  \cite{hendrickx2024sweet, bassi2024optimaloperationholespin, wang2024operating}. Coherence will likely further improve with the use of isotopically purified heterostructures. 

 Non-resonant effects, arising from diabatic pulsing of $J$ (\supref{app:basebandSWAP}) and breaking of the rotating wave approximation, can become limiting for reaching higher fidelities; however, this can likely be mitigated by synchronization and pulse shaping \cite{Philips2023}, as demonstrated for other resonant gate schemes \cite{noiri2022fast, zwanenburg2025singlequbitgatesrotatingwaveapproximation}. Both the drive and ramp durations can be reduced with a larger \(\Delta E_\mathrm{Z}\) while ensuring compliance
with the rotating wave approximation and adiabaticity ($J < \Delta E_\mr{Z}$). In this experiment this value was only $\Delta E_\mathrm{Z}\approx0.12 \overline{E}_\mathrm{Z}$. However, values of $\Delta E_\mathrm{Z}>0.5 \overline{E}_\mathrm{Z}$ are regularly observed in Ge/SiGe devices for similar magnetic fields \cite{wang2024operating,hendrickx2021four}, which would allow for significant gate speedup, thereby reducing the impact of dephasing.

In summary, we have demonstrated methods for tuning the parameters of a resonant fSim gate in spin qubits. This enabled the benchmarking of an iSWAP gate between two Loss-Divincenzo qubits, yielding a maximum fidelity of 93.8(5)\% in interleaved randomized benchmarking. Our implementation supports concurrent drive and baseband pulses as well as ramps, making it compatible with efficient resonant gate schemes \cite{Philips2023}.  

Our results establish a practical path towards realizing the highly efficient fermionic simulation two-qubit gate in spin qubits. The demonstrated advantages of the fSim gate in various quantum algorithms \cite{arute2019quantum, arute2020observationseparateddynamicscharge, Yu2025, Kivlichan2018, Verstraete2009, iSwap2003} further highlight the potential of spin qubits. Beyond computation, the native implementation of fSim opens new possibilities for quantum simulations, enabling more efficient modeling of molecular electronic structures and strongly correlated fermionic systems with spin qubits.

\begin{acknowledgments}
\paragraph*{\bf Funding}
This project acknowledges funding from the European Union's Horizon 2020 research and innovation programme under the Marie Skłodowska-Curie grant agreement no.\ 847471. In addition, this project has received funding from the NCCR SPIN under grant no.\ 51NF40-180604 and 51NF40-225153, and under the grant no.\ 200021-188752 of the Swiss National Science Foundation. We also thank Michael Stiefel and all the Cleanroom Operations Team of the Binnig and Rohrer Nanotechnology Center (BRNC) for their help and support.
\paragraph*{\bf Author contributions}
KT and AO conceived and performed the experiment. 
BH and AO developed the theory model. 
BH and JA evaluated the quantum process tomography data. 
FJS and MM fabricated the device. UvL contributed to the measurements of the device. 
LS, IS, KT, FJS, PHC and MM developed parts of the device. 
SB grew the heterostructure. 
GS, IS and  PHC contributed to the development of the measurement software. 
SP, MA, EGK and GS contributed to the development of the experimental setup. 
MPV and LM contributed to the interpretation of the measurements. 
KT, AO, BH and PHC wrote the manuscript with input from all authors. 
GS, AF and PHC supervised the project.
\paragraph*{\bf Competing interests}
The authors declare no competing interests.
\paragraph*{\bf Data availability}
The data and analysis that support the findings of this study will be available in a Zenodo repository. 
\end{acknowledgments}

\bibliographystyle{apsrev4-1-title} 
\bibliography{refs.bib}
\clearpage

\onecolumngrid 
\clearpage
{\centering
\large\textbf
{Supplementary information for: \\ Resonant two-qubit gates for fermionic simulations with spin qubits} \\ \rule{0pt}{12pt}
}

\newcommand{\RefFigResonant}{\figref{fig:resonant}} 
\newcommand{\RefFigDipersivePhi}{\figref{fig:dispersivephi}} 
\newcommand{\RefFigDipersivePhotonNumber}{\figref{fig:dispersivephotonnumber}} 

\appendix

\section{Coherence times}
\label{app:Coherence times}

\begin{figure}[h!]
    \centering
    \includegraphics[width=8 cm]{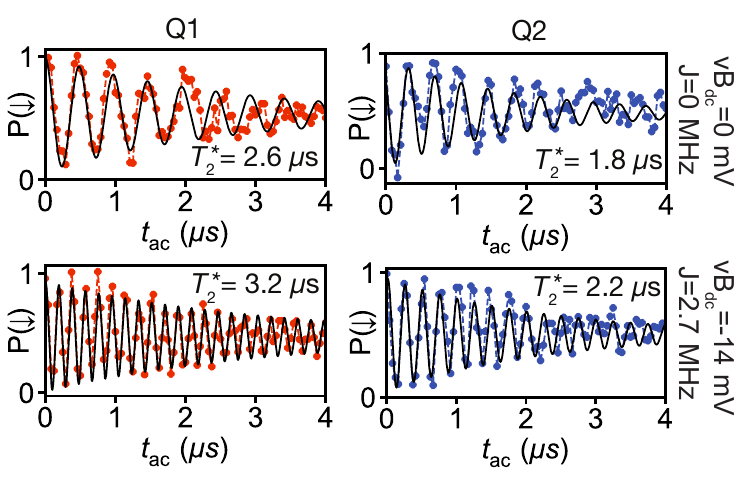}
    \caption{The free induction decay obtained from Ramsey experiments for Q1 (left) and Q2 (right) at $\mr{vB}_\mathrm{\text{dc}}=0$ (upper) and $\mr{vB}_\mathrm{\text{dc}}=-14$ (lower). The coherence times $T_{2}^*$ for all four measurements are shown inside the plots. }
    \label{fig:A1}
\end{figure}

\section{Baseband SWAP Oscillations}
\label{app:basebandSWAP}

\begin{figure*}[tbp]
\includegraphics[width=14cm]{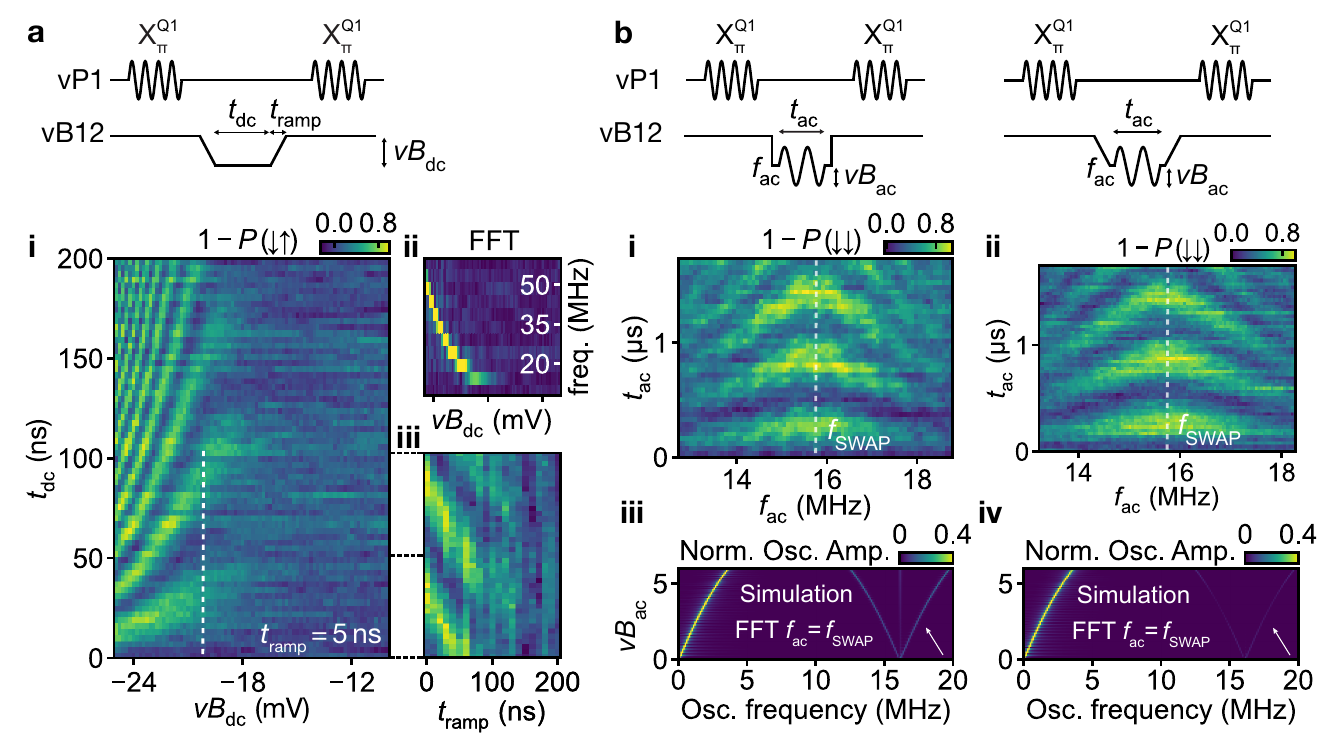}
\caption{\textbf{Exchange oscillations between two neighboring spins} \textbf{(a)} (i) Oscillations induced in the population of the $|{\uparrow\downarrow}\rangle$ state as a function of the amplitude $\mr{vB}_\mr{\text{dc}}$ of a baseband square pulse applied on the barrier. (ii) Fast Fourier transform of (i) revealing the oscillation speed. (iii) Monitoring the $|{\uparrow\downarrow}\rangle$ oscillation upon introducing a Tukey-shaped ramp to the barrier pulse. \textbf{(b)} Coherent $|{\uparrow\downarrow}\rangle$ $\leftrightarrow$ $|{\downarrow\uparrow}\rangle$ oscillations induced by superposing an oscillating signal of amplitude $\mr{vB}_\mr{\text{ac}}$ with a baseband barrier pulse. A ramp time of $t_\mr{ramp} = 5 $\,ns and $t_\mr{ramp} = 100$\,ns was chosen for (i) and (ii), respectively. For the same ramp times, (iii) and (iv) show the FFT of the simulated $|{\uparrow\downarrow}\rangle$ $\leftrightarrow$ $|{\downarrow\uparrow}\rangle$ oscillation for a drive frequency of $f_\mr{\text{ac}}=f_\mr{SWAP}$. The color-scale has been normalized between 0 and 1 and is plotted until 0.4 to increase the visibility of the higher frequency oscillations (white arrow).  }
\label{fig:fig2}
\end{figure*}

A rapid increase of $J$ results not only in conditional frequency shifts of the spins but can also induce exchange oscillations between the two antipolarized spin states \cite{ni2025swap, petit2022design}. This occurs because the two-spin system eigenstates change diabatically from predominantly antipolarized, for $J=0$, to predominantly singlet-triplet-like, for $J>\Delta E_z$. 

After initializing the spins in the $|{\uparrow\downarrow}\rangle$ state and introducing again a square voltage pulse on the barrier of amplitude $\mr{vB}_\mr{\text{dc}}$, we observe the $|{\uparrow\downarrow}\rangle$ and $|{\downarrow\uparrow}\rangle$ to undergo swap oscillations. The speed of this oscillation reaches 60 MHz for $\mr{vB}_\mr{\text{dc}}=-25$\,mV as seen in the fast Fourier transform (FFT) of \figref{fig:fig2}(a).
The shape of the vB12 pulse is crucial for this operation, as a slow enough ramp can result in an adiabatic transition and hence no oscillations. To demonstrate this, we incorporate a Tukey ramp of variable duration to a barrier pulse of amplitude $\mr{vB}_\mr{\text{dc}}=-20$ mV. As the ramp time increases from 0 to 200\,ns, the oscillations gradually disappear (see \figref{fig:fig2}(a)(iii)) illustrating the transition from the diabatic to the adiabatic regime. While the diabatic process has been utilized to perform baseband SWAP gates, it relies on achieving large $J$ or composite rotations, making the qubits more susceptible to charge noise. 


The two experiments of resonant exchange driving swap oscillations shown in Fig.\,2b(i,ii) differ in the ramp time of the dc barrier pulse, with $t_\mr{ramp} = 5 $\,ns and $t_\mr{ramp} = 100$\,ns, respectively, as can be seen in the schematics of the pulse sequences. We refer to the first case as diabatic and the second as adiabatic. For the diabatic case, secondary fast oscillations are superimposed with the resonant exchange oscillation, which are suppressed in the adiabatic case. To understand the origin of these oscillations we perform a unitary simulation at $f_\mathrm{ac}=f_\mathrm{SWAP}$, for the corresponding ramp times, varying the drive amplitude $\mr{vB}_\mr{\text{ac}}$. In \figref{fig:fig2}(b)(iii,iv) we show the fast Fourier transform of the results. Firstly, we observe a main component originating at zero frequency revealing the nonlinear dependence of the resonant SWAP speed (Rabi frequency) on the drive amplitude. The intensity of this curve, reflecting the amplitude of the driven oscillation, is unchanged between the two ramp times. Additionally, for $t_\mr{ramp}$\,=\,5\,ns, we observe the appearance of three additional weak frequency components originating at the baseband SWAP frequency ($\Delta \widetilde{E}_z$) at zero drive amplitude. The intensity of these curves is reduced significantly for $t_\mr{ramp}$\,=\,100\,ns confirming their diabatic baseband SWAP origin. As for the nature of these frequencies, while the central one corresponds to the baseband swap oscillation with frequency $\Delta\widetilde{E}_{\mathrm{z}}$, the two side-bands are attributed to a dressing of the baseband swap oscillation by the resonant drive, similar to Refs.\ \cite{laucht2017dressed,tsoukalas2025dressed}.

\section{Unitary evolution and single qubit corrections}
\label{app:Unitary evolution}

Here we relate the parameters of the fSim (defined as $\mr{fSim}(\gamma, \zeta) = \mr{iSWAP}^{-2\gamma/\pi}\mr{CZ}^{\zeta/\pi}$) gate to the experimentally measurable quantities, by deriving the time evolution according to the time-dependent Schr\"odinger equation when the baseband pulse is applied. Throughout this section we will write the unitary gates in the frame where each qubit is rotating with its respective Larmor frequency.

\begin{figure*}[tbp]
    \centering
    \includegraphics[width=0.7\textwidth]{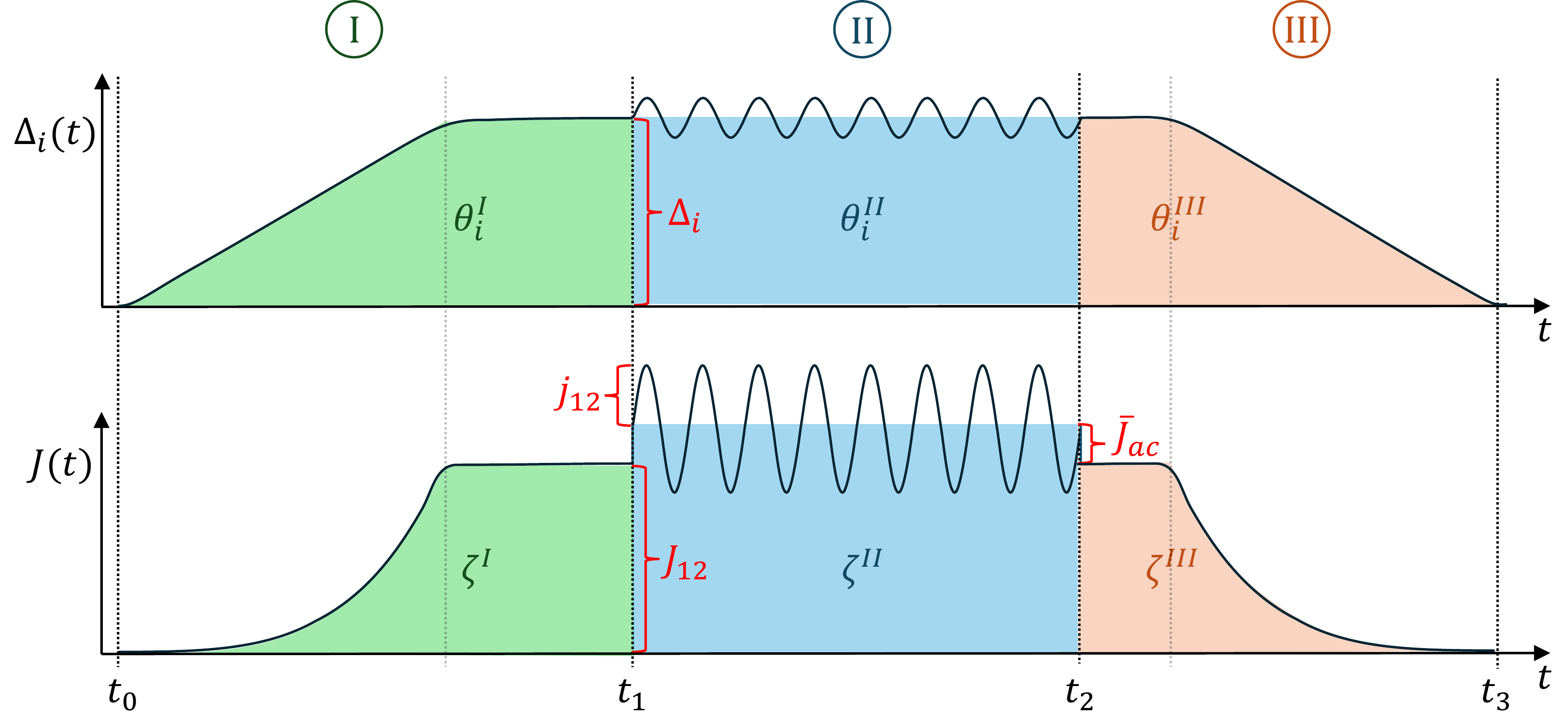}
    \caption{Parameters of the time dependent Hamiltonian when the baseband pulse is applied. For the driven part of the exchange interaction, we only plot the sum of the first two Fourier components that are relevant in the rotating-wave approximation, i.e., $J(t) = J_\mathrm{dc}(t) + \Theta (t-t_1)\Theta(t_2-t)(\bar{J}_\mathrm{ac} +j_{12}\sin(\omega t + \phi))$.}
    \label{fig:baseband_SM}
\end{figure*}

We divide the total time evolution into three intervals as shown in \figref{fig:baseband_SM}, each with a corresponding propagator. The gate could then be constructed as
\begin{eqnarray}
    \mathrm{fSim}(\gamma, \zeta) = U_\mathrm{post}U^{III}U^{II}U^{I}U_\mathrm{pre},
\end{eqnarray}
where $U^{I}$, $U^{II}$, and $U^{III}$ are the corresponding parts of the time-evolution, while $U_\mathrm{pre}$ is the single qubit phase correction applied before the baseband pulse to correct the phases of interval I, and $U_\mathrm{post}$ contains phase corrections for intervals II and III. We will show that all phase corrections can instead be pushed to the end of the baseband pulse, i.e.,
\begin{eqnarray}
    \mathrm{fSim}(\gamma, \zeta) = U_\mathrm{post}'U^{III}(U^{II})'U^{I},
\end{eqnarray}
where $(U^{II})'$ only differs from $U^{II}$ by a shift of the drive phase and the single phase-correction step can be realized using virtual Z gates.

\subsection{Time evolution during interval I}

During the first part of the baseband pulse, the two-qubit system evolves according to the Hamiltonian
\begin{equation}
    H^I(t)=\frac{hf_1+ h\Delta_1(t)}{2} \sigma_z\!\otimes  \mathbb{1} +\frac{hf_2+ h\Delta_2(t)}{2} \mathbb 1 \otimes \sigma_z + 
    \frac{1}{4}J_\mathrm{dc}(t) \sum_i \sigma_i \otimes \sigma_i,
\end{equation}
where $\Delta_i$ is the difference between the Larmor frequency of qubit $i$ at time $t$ and its frequency without the baseband pulse. We calculate the unitary time-evolution (in the rotating frame) according to the Hamiltonian above as $U = \text{exp}{(i\pi f_1t \sigma_z\!\otimes  \mathbb{1} + i\pi f_2t\mathbb{1} \otimes \sigma_z)} \text{exp}{[-\frac i \hbar \int \dd tH(t)]}$ which leads to
\begin{equation}
    U^I=\begin{pmatrix}
        e^{-\frac i 2(\theta_1^I+\theta_2^I+\zeta^I)} &0&0&0\\
        0& e^{-\frac i 2(\theta_1^I-\theta_2^I)} &0&0\\
        0&0& e^{\frac i 2(\theta_1^I-\theta_2^I)} &0\\
        0&0&0& e^{\frac i 2(\theta_1^I+\theta_2^I-\zeta^I)}\\
    \end{pmatrix},
\label{eq:UI}
\end{equation}
where the corresponding angles are
\begin{eqnarray}
    \theta^I_i = 2\pi \int \limits_{t_0}^{t_1} \!\dd t\, \Delta_i(t),
    \label{eq:thetaI}
    \\
    \zeta^{I} = \frac 1 \hbar \int \limits_{t_0}^{t_1} \!\dd t\, J_\mathrm{dc}(t).
    \label{eq:zetaI}
\end{eqnarray}
It is important to note that \eqnref{eq:UI} is valid in two limiting cases {\it (i)} if $J_\mathrm{dc}\ll hf_1-hf_2$ where the swap oscillations with amplitude $J_\mathrm{dc}/(hf_1-hf_2)$ are negligible; {\it (ii)} in the adiabatic ramp case where Eqs.~\eqref{eq:thetaI}-\eqref{eq:zetaI} need to be calculated from the instantaneous eigenenergies as
\begin{eqnarray}
    \theta^I_\mathrm{1,ad} &=& \frac 1 {2\hbar} \int \limits_{t_0}^{t_1} \!\dd t\, [E_{\uparrow\uparrow}(t)-E_{\downarrow\downarrow}(t)+E_{\uparrow\downarrow}(t)-E_{\downarrow\uparrow}(t)],
    \\
    \theta^I_\mathrm{2,ad} &=& \frac 1 {2\hbar} \int \limits_{t_0}^{t_1} \!\dd t\, [E_{\uparrow\uparrow}(t)-E_{\downarrow\downarrow}(t)-E_{\uparrow\downarrow}(t)+E_{\downarrow\uparrow}(t)],
    \\
    \zeta^{I}_\mathrm{ad} &=& \frac 1 {\hbar} \int \limits_{t_0}^{t_1} \!\dd t\, [E_{\uparrow\uparrow}(t)+E_{\downarrow\downarrow}(t)].
\end{eqnarray}

In order to convert this unitary to a controlled phase type contribution of the fSim gate, we apply the single-qubit phase corrections
\begin{eqnarray}
    U_\mathrm{pre} = Z_1(\theta_1^I+ \tfrac 1 2\zeta^I)\,Z_2(\theta_2^I + \tfrac 1 2\zeta^I),
\label{eq:Upre}
\end{eqnarray}
where $Z_1(\varphi) = \exp(i\frac \varphi 2\sigma_z\otimes \mathbb{1})$, and $ U^I U_\mathrm{pre} = U_\mathrm{pre} U^I = \mathrm{CPHASE}(-\zeta^I)$. We note that $U_\mathrm{pre}$ could be applied as a virtual correction before the baseband pulse, provided that the phase of the drive is also updated according to the corrected phase difference of the two qubits. In this work, however, we apply the single qubit phase corrections in a single step after the two-qubit gate.

\subsection{Time evolution during interval II}

In the second time interval the system evolves according to
\begin{equation}
    H^{II}(t)=\frac{hf_1+ h\Delta_1}{2} \sigma_z\!\otimes  \mathbb{1} +\frac{hf_2+ h\Delta_2}{2} \mathbb 1 \otimes \sigma_z + 
    \left(J_{12} + \bar{J}_\mathrm{ac} + j_{\mathrm{12}} \sin (2\pi f_\mathrm{SWAP} (t-t_1) + \phi)\right) \sum_i \sigma_i \otimes \sigma_i
\end{equation}
where the resonance condition is $f_\mathrm{SWAP} = f_1+ \Delta_1 - (f_2+ \Delta_2)$ and $\phi$ is the phase in which the drive starts at $t=t_1$ in the rotating frame. I.e., the calculated value of $\phi$ below is directly applicable if {\it (i)} the relative phase of the qubits is zero at $t=0$ and {\it (ii)} the phase of the drive is continuously updated using the frequency $f_1-f_2$. Solving this Hamiltonian in the rotating-wave approximation leads to
\begin{eqnarray}
    U^{II} = \begin{pmatrix}
        e^{-\frac i 2 (\theta_1^{II}+\theta_2^{II}+\zeta^{II})} &0&0&0\\
        0& e^{-\frac i 2(\theta_1^{II}-\theta_2^{II})}\cos (\frac{ j_{12}t_\mathrm{ac}}{4\hbar}) & e^{-\frac i 2(\theta_1^{II}-\theta_2^{II})}e^{-i\phi}\sin (\frac{ j_{12}t_\mathrm{ac}}{4\hbar})&0\\
        0&-e^{\frac i 2(\theta_1^{II}-\theta_2^{II})} e^{i\phi}\sin (\frac{j_{12}t_\mathrm{ac}}{4\hbar})& e^{\frac i 2(\theta_1^{II}-\theta_2^{II})}\cos (\frac{j_{12}t_\mathrm{ac}}{4\hbar}) &0\\
        0&0&0& e^{\frac i 2(\theta_1^{II}+\theta_2^{II}-\zeta^{II})}\\
    \end{pmatrix},
\label{eq:UII}
\end{eqnarray}
where $t_\mathrm{ac} = t_2-t_1$ and we define the first argument of the fSim gate as $\gamma = \frac{j_{12}t_\mathrm{ac}}{4\hbar}$. Furthermore
\begin{eqnarray}
    \theta^{II}_i &=& 2\pi \Delta_i t_\mathrm{ac},\\
    \zeta^{II} &=& \frac 1 \hbar (J_{12} + \bar{J}_\mathrm{ac})t_\mathrm{ac}.
\end{eqnarray}
In order to get rid of the single-qubit phases from \eqnref{eq:UII} we need to apply the phase-correction
\begin{eqnarray}
    U_\mathrm{post}^{II} = Z_1(\theta_1^{II} + \tfrac 1 2\zeta^{II})\,Z_2(\theta_2^{II}+\tfrac 1 2\zeta^{II}),
\end{eqnarray}
after the time evolution.

Now we bring the single-qubit phase corrections from \eqnref{eq:Upre} across $U^{II}$ while allowing for changes in the controllable parameters of the Hamiltonian
\begin{eqnarray}
    U^{II}U_\mathrm{pre} = U_\mathrm{pre} U^{II}(\phi\rightarrow \phi + (\theta_1^I - \theta_2^I)).
\end{eqnarray}

{\it Special case:} For the iSWAP gate ($\gamma=\pi/2$), instead of modifying the drive phase, we can adjust the single-qubit corrections to achieve
\begin{eqnarray}
    U^{II}(\gamma=\pi/2)U_\mathrm{pre} = U_\mathrm{pre,iSWAP} U^{II}(\gamma=\pi/2).
\end{eqnarray}
where the modified single-qubit corrections are
\begin{eqnarray}
    U_\mathrm{pre,iSWAP} = Z_1(\theta_2^I+ \tfrac 1 2\zeta^I + \phi + \pi/2)\,Z_2(\theta_1^I + \tfrac 1 2\zeta^I - \phi - \pi/2).
\end{eqnarray}
Note that we the single qubit corrections for the first section are exchanged, and we have added (subtracted) $\phi+\pi/2$ to the phase of the first (second) qubit.

\subsection{Time evolution during interval III}

The calculation in the third time interval is analogous to that of I. Therefore we just write the correction required to convert this section into a $\mathrm{CPHASE}(-\zeta^{III})$ gate:
\begin{eqnarray}
    U_\mathrm{post}^{III} = Z_1(\theta_1^{III} + \tfrac 1 2\zeta^{III})\,Z_2(\theta_2^{III}+\tfrac 1 2\zeta^{III}),
\end{eqnarray}
where
\begin{eqnarray}
    \theta^{III}_i = 2\pi \int \limits_{t_2}^{t_3} \!\dd t\, \Delta_i(t),\\
    \zeta^{III} = \frac 1 \hbar \int \limits_{t_2}^{t_3} \!\dd t\, J_\mathrm{dc}(t).
\end{eqnarray}

\subsection{Gate parameters and corrections}

Finally we exploit that $U^{III}$ commutes with the single qubit phase corrections that are required in interval I and II, and express the arguments of the fSim gate as 
\begin{eqnarray}
    &\gamma = \frac{j_{12}t_\mathrm{ac}}{4\hbar}&\\
    &\zeta = \frac 1 \hbar \int \limits_{t_0}^{t_3} \!\dd t\, J_\mathrm{dc}(t) + \frac 1 \hbar \bar{J}_\mathrm{ac}t_\mathrm{ac}.&
\end{eqnarray}
The single qubit corrections, to be applied after the baseband pulse, are
\begin{eqnarray}
    U_\mathrm{post}' &=& Z_1(\theta_1+ \tfrac 1 2\zeta)\,Z_2(\theta_2 + \tfrac 1 2\zeta),\\
    \theta_i &=& 2\pi \int \limits_{t_0}^{t_3} \!\dd t\, \Delta_i(t).
\end{eqnarray}
Moreover, the starting phase of the drive needs to be $\phi = \pi/2 + (\theta_1^I - \theta_2^I)$ (wrt.\ the relative phase of the qubits without the baseband pulse) to recover the canonical form of the fSim gate.

{\it Special case:} For the iSWAP gate ($\gamma=\pi/2$ and $\zeta = 2\pi$), instead of modifying the drive phase, we can adjust the single-qubit corrections as
\begin{eqnarray}
    U'_\mathrm{post} = Z_1(\theta_2^I + \theta_1^{II} + \theta_1^{III} + \zeta/2 + \phi + \pi/2)\,Z_2(\theta_1^I + \theta_2^{II} + \theta_2^{III} + \zeta/2 - \phi - \pi/2),
\label{eq.UpostiSWAP}
\end{eqnarray}
where $\phi$ is an arbitrary phase in which the drive has started. Finally, we define the residual phase as $\phi_{\mathrm{res}} = -\phi -\pi/2 + (\theta_{1}^I - \theta_{2}^I)$. When this is substituted in Eq.~\ref{eq.UpostiSWAP} we arrive at 
\begin{eqnarray}
    U'_\mathrm{post} = Z_1(\theta_1 + \zeta/2 - \phi_{res})\,Z_2(\theta_2 + \zeta/2 + \phi_{res}).
\label{eq.phi_res}
\end{eqnarray}

\subsection{Resonant drive oscillator tracking}
\label{app:resdrive}

In order to keep track of the phase for the resonant drive we make use of a separate oscillator that evolves at the swap drive frequency $f_{\mathrm{SWAP}}$. In order to synchronize this oscillator with the difference-frame of the idle qubits, we add a phase $\widetilde{\phi} = 2\pi [(f_\mathrm{1} - f_\mathrm{2}) - f_{\mathrm{SWAP}}]t_{\mathrm{abs}}$ to the oscillator phase at the start of the resonant drive, where $t_{\mathrm{abs}}$ is the time from the start of the experiment and $f_\mathrm{1}, f_\mathrm{2}$ are the idle qubit frequencies. 
This is particularly relevant to our work, where the drive of the iSWAP gate is facilitated by an additional AWG oscillator and not by subtracting the drive signals of the two single spin oscillators as in other works \cite{Sigillito2019}.
 
Whenever a virtual ($R_z$) gate is applied to a single qubit this constitutes a re-definition of the reference frame for this qubit, shifting the phase of the oscillator for all subsequent gates. This therefore also re-defines the difference-frame of the two-qubit system. A phase is therefore added (subtracted) to (from) the resonant drive for virtual gates on qubit 1 (qubit 2). We note that the compatibility of this gate implementation with virtual Z rotations renders it suitable for generating a highly general set of continuous gates beyond the fSim gate, such as for instance Givens rotations applied in quantum chemistry simulations \cite{arute2020observationseparateddynamicscharge}. 
 
\subsection{Simulation}
\label{app:drivesim}

The simulations in \figref{fig:fig2}(b) were done with the Hamiltonian in the $\{\ket{\uparrow\downarrow},\ket{\downarrow\uparrow}\}$ basis: 
\begin{equation*}
    H = \begin{pmatrix}
-\Delta E_z & J(t)/2 \\
J(t)/2 & \Delta E_z 
\end{pmatrix}	
\end{equation*}
where $J(t)$ is extracted from the measured dependence of $J$ on $\mr{vB}_{\text{dc}}$.

\section{Gate tuning}
\label{app:Gatetuning}

The general formula for conditional oscillations with Q1 being the target qubit is:
\begin{align*}
    P_{Q1, Q2\uparrow} &= \frac{1}{2} \left(1 - \cos(2\pi\int_{t_\mathrm{tot}} [J(t) /2h + \Delta_1(t)]\dd t + \theta   )\right) \\
        P_{Q1, Q2\downarrow} &= \frac{1}{2} \left(1 - \cos(2\pi\int_{t_\mathrm{tot}} [J(t)/2h + \Delta_1(t)]\dd t - \theta  )\right) 
\end{align*}
where $\theta$ is an added phase at the end of the Ramsey experiment. Analogously for Q2. In the case of negligible $t_{\mathrm{ramp}}$, as for the case $t_{\text{ramp}} = 5$\,ns, the exchange energy can be extracted as the frequency of Ramsey oscillations as a function of $t_{\mathrm{dc}}$. 

\section{Process tomography}
\label{app:proc_tomo}

\subsection{Notation}

For the calculations of this section we adapt the super-Dirac notation, where linear operators $O$ acting on the Hilbert space are represented as $d^2$ dimensional vectors $\sket{O}$ \cite{Nielsen2021gatesettomography}. Since linear operators form a vector space, we can write any operator $\sket{O}$ as
\begin{eqnarray}
    \sket{O} = \sum \limits_{i=0}^{d^2} c_i \sket{B_i},
    \label{eq:lincomb}
\end{eqnarray}
provided that the basis operators $\sket{B_i}$ are linearly independent. In the context of tomography, such a basis is called informationally complete (IC). For a tomography measurement it is useful to find an IC basis in terms of pure states that can be easily prepared and measured. Forming an IC basis from pure states requires that the determinant of the $d^2\times d^2$ matrix $[\sket{B_1},\dots,\sket{B_{d^2}}]$ is nonzero. For $n$ qubits such an IC basis is provided by the set of operators $\{B_i\} = \{\ket 0 \bra 0\!,\,\ket 1 \bra 1\!,\,\ket + \bra +\!,\,\ket{+i} \bra{+i}\}^{\otimes n}$.

In this framework the scalar product corresponds to the trace of the product of operators, i.e., 
\begin{eqnarray}
    &\llangle O_1\sket{O_2} \equiv {\Tr} [O_1^\dagger O_2] = \sum \limits_{ij} (O_1)^*_{ij} (O_2)_{ji}.
\end{eqnarray}
If the basis is orthonormal, that is IC and $\llangle B_i \sket{B_j} = \delta_{ij}$, the coefficients $c_i$ from \eqnref{eq:lincomb} can be found as $c_i = \llangle B_i \sket{O}$. The IC basis of pure states is not an orthonormal basis, therefore in the following we will use the basis of ($n$-qubit) Pauli matrices, i.e., $\{B_i\} = \{2^{-n/2} P_i\}$.

Quantum channels are defined as linear maps over the vector space of operators. Physical channels $\Lambda$ are completely positive trace-preserving (CPTP) maps, i.e., $\Lambda \in \mathcal O_\mathrm{CPTP}$. The channel $\Lambda$ acts on the density matrix $\rho$ as follows
\begin{eqnarray}
    &\rho'_{kl} = \Lambda[\rho]_{kl} = \sum \limits_{kl} \Lambda_{ij,kl} \rho_{kl}\\
    &\sket{\rho'} \equiv \Lambda \sket{\rho}.
\end{eqnarray}
Quantum channels can also be expressed using an orthonormal set of operators, e.g.,
\begin{eqnarray}
    \Lambda = \sum \limits_{i=0}^{d^2} R_{ij} \sket{P_i} \sbra{P_j},
\end{eqnarray}
where $R_{ij}$ is the Pauli transfer matrix (PTM), whose elements are shown in \figref{fig:fig4}. The first row of the PTM is $R_{0j} = \delta_{0j}$ due to trace preservation. Furthermore, the first column $R_{i0}$ for $i\neq0$ maps maximally mixed states ($\rho \propto P_0$) into nontrivial states, which can be interpreted as a relaxation-type process.

\subsection{Process tomography}

In process tomography, we prepare fiducial states, act on them with the studied gate or channel and perform a set of fiducial measurements on the state \cite{altepeter2003ancilla,artiles2004invitation}. Fiducial measurements on the state of a single qubit are $X$, $Y$, and $Z$ measurements that are informationally complete. Here we use an overcomplete set of fiducial $n$-qubit states
\begin{eqnarray}
    \{ \sket{\rho_0}, \sket{\rho_1}, \sket{\rho_+}, \sket{\rho_-}, \sket{\rho_{+i}}, \sket{\rho_{-i}} \}^{\otimes n},
    \label{eq:fid_states}
\end{eqnarray}
where $\sket{\rho_\alpha}$ is a the superket notation of the pure state $\rho_\alpha = \ket \alpha \bra \alpha$. Fiducial measurements on $n$ qubits are the $3^n$ measurements on the basis
\begin{eqnarray}
    \{ X,Y,Z \}^{\otimes n},
\end{eqnarray}
each measurement can have two possible outcomes, leading to $6^n$ projectors
\begin{eqnarray}
    \{ \sbra{\rho_0}, \sbra{\rho_1}, \sbra{\rho_+}, \sbra{\rho_-}, \sbra{\rho_{+i}}, \sbra{\rho_{-i}} \}^{\otimes n},
\end{eqnarray}
where $\sbra{\rho_\alpha}$ is a the superbra notation of the projector $\rho_\alpha = \ket \alpha \bra \alpha$. The measurement expectation values of tomography circuits performed on a channel $\Lambda$ are
\begin{eqnarray}
    P_{ij} = \sbra{\rho_i} \Lambda \sket{\rho_j}.
\end{eqnarray}
The channel can be recovered from this matrix as 
\begin{eqnarray}
    \Lambda = (A^\dagger)^{-1} P A^{-1},
    \label{eq:rec_chan}
\end{eqnarray}
where $A$ is a $4^n\times 6^n$ matrix formed by the fiducial states.

Importantly, state preparations and measurements are always done using the $\ket{\downarrow\downarrow}$ state and applying some $\pi/2$ or $\pi$ pulse gates to prepare the fiducial basis. The former process, however, is hindered by SPAM and single-qubit gate errors. Since single qubit gate errors are generally much lower than two-qubit gate errors (e.g., they are used in the calibration of two-qubit gates) we neglect their contribution and write the outcome matrix as
\begin{eqnarray}
    P^G_{i,j} = \sbra{E_{\downarrow\downarrow}} \Lambda_m \mathcal U_i^\dagger \Lambda_G \mathcal U_j \Lambda_p \sket{\rho_{\downarrow\downarrow}} .
    \label{eq:PG_SPAM}
\end{eqnarray}
where $\mathcal U_i$ is an ideal unitary channel defined by $\sket{\rho_i} = \mathcal U_i\sket{\rho_{\downarrow\downarrow}}$, $\Lambda_{m(p)}$ is the channel of measurement (preparation) errors and $\Lambda_G$ is the two-qubit gate to be characterized.

If we tried to recover a channel from \eqnref{eq:PG_SPAM} using \eqnref{eq:rec_chan}, we would find that the channel is not physical due to the presence of SPAM errors ($\Lambda \notin \mathcal O_\mathrm{CPTP}$). This is a common problem in tomography methods, including GST, but SPAM errors can be mitigated to some degree by making assumptions on the channels $\Lambda_{m(p)}$. We assume that the SPAM channels commute with the single qubit Clifford gates $\mathcal U_i$ to get
\begin{eqnarray}
    P^G_{i,j} = \sbra{E_i} \Lambda_m \Lambda_G \Lambda_p \sket{\rho_j} ,
    \label{eq:PG_SPAM2}
\end{eqnarray}
where the assumption holds exactly for single and two-qubit depolarizing channels. In general, however, it is not possible to find a single physical channel $\tilde \Lambda_p$ such that $\mathcal U_j \Lambda_p \sket{\rho_{\downarrow\downarrow}} = \tilde \Lambda_p \sket{\rho_j}$ \cite{blume2024easy}. 

In order to obtain the quantum channel of the full process, we extract a channel from \eqnref{eq:PG_SPAM2} as 
\begin{eqnarray}
    \tilde\Lambda_\mathrm{full} = (A^\dagger)^{-1} P A^{-1}.
    \label{eq:Lambdafull}
\end{eqnarray}
Due to shot noise and the violation of our assumptions on the SPAM channels the channel $\tilde\Lambda$ obtained this way is not yet physical. To ensure that the resulting channel is physical, we have to solve an optimization problem. We use three different methods.

{\it Standard method.} The standard method simply projects the channel resulting from \eqnref{eq:Lambdafull} onto the closest physical channel such that
\begin{equation}
    \Lambda_\mathrm{full} = \mathrm{argmin}_{\Lambda} ||\tilde\Lambda_\mathrm{full}-\Lambda||_{F}\, |\,
    \Lambda \in \mathcal O_\mathrm{CPTP},
    \label{eq:CPTP}
\end{equation}
where $||\Lambda||_{F} = \sqrt{\Tr[\Lambda^{\dagger}\Lambda]}$ is the Frobenius distance. The optimization problem above can always be solved efficiently because of the convexity of the function \cite{diamond2016cvxpy}.

{\it Least-square method.} The least-square method minimizes the $l_2$ norm of the deviation between the measured outcomes and the parametrized outcome where the latter is subject to the physical constraint. I.e.,
\begin{equation}
    \Lambda_\mathrm{full} = \mathrm{argmin}_{\Lambda} ||P - A^\dagger \Lambda A||_2\, |\,
    \Lambda \in \mathcal O_\mathrm{CPTP}.
\end{equation}
The main difference between the least-square and the standard method is that we enforce the CPTP constraint during the optimization in the least-square case. The optimization is done using gradient decent (10'000 steps, learning rate 0.01), and the CPTP constraint is enforced at every step by solving \eqnref{eq:CPTP} using {\texttt CVXPY} \cite{diamond2016cvxpy}.

{\it Maximum-likelihood method.} In the case of the maximum-likelihood (ML) method the cost function to be minimized is 
\begin{equation}
    \Lambda_\mathrm{full} = \mathrm{argmin}_{\Lambda} \left[-\sum \limits_{i,j} P_{ij} \log( A_i^\dagger \Lambda A_j)\right]\, |\,
    \Lambda \in \mathcal O_\mathrm{CPTP}.
\end{equation}
Similarly, the optimization is done using gradient decent (1'000 steps, learning rate 0.01), and the CPTP constraint is enforced at every step by solving \eqnref{eq:CPTP} using \texttt{CVXPY} \cite{diamond2016cvxpy}.

We can calculate the process fidelity of the physical channel $\Lambda_\mathrm{full}$ with the perfect iSWAP gate. See \tabref{tab:fidelity_compare} for the result obtained from different methods.
For the process fidelity of channel $\Lambda$ with respect to the ideal iSWAP gate, is defined as
\begin{equation}
    \mathcal{F}_\mathrm{p} (\Lambda,\mathcal U_\mathrm{iSWAP}) = \frac 1 {d^2} \Tr \left[\Lambda\,  \mathcal U_\mathrm{iSWAP} \right]^2,
\end{equation}
where $\mathcal U_\mathrm{iSWAP}$ is the channel corresponding to the ideal iSWAP gate. Process fidelities with a single argument $\mathcal{F}_\mathrm{p} (\Lambda)$ hereinafter refer to the process fidelity of $\Lambda$ with the ideal iSWAP gate. The average gate fidelity can then be calculated as
\begin{equation}
    \mathcal{F}_\mathrm{g}(\Lambda) = \frac{\mathcal{F}_\mathrm{p}(\Lambda)d+1}{d+1}.
\end{equation}

\begin{table}
	\centering
    \caption{Comparison of different optimization methods for the process fidelity $\mathcal{F}(\Lambda_\mathrm{full})$ and gate fidelity $\mathcal{F}_\mathrm{g}(\Lambda_{-p})$ ($\mathcal{F}_\mathrm{g}(\Lambda_{-m})$) of the iSWAP gate, where SPAM errors are removed assuming they are purely preparation (measurement errors). Process fidelity of the identity gate shows the effect of SPAM errors.}
    \label{tab:fidelity_compare}
    \begin{tabular}{cccc}
    \multicolumn{4}{c}{iSWAP (adiabatic)}
    \\
    \hline \hline
    Fidelity & Standard & Least-square & ML \\ 
    \hline 
    $\mathcal{F}_\mathrm{p}(\Lambda_\mathrm{full})$  & 61.9\% & 62.7\% & 59.3\%\\ 
    $\mathcal{F}_\mathrm{g}(\Lambda_{-p})$  & 84.4\% & 83.9\% & 81.7\% \\
    $\mathcal{F}_\mathrm{g}(\Lambda_{-m})$ & 83.3\% & 83.2\% & 80.8\%\\ 
    \hline \hline
    \end{tabular}
    \\\vspace{0.2cm}
    \begin{tabular}{cccc} 
    \multicolumn{4}{c}{iSWAP (diabatic)}
    \\
    \hline \hline
    Fidelity & Standard & Least-square & ML \\ 
    \hline 
    $\mathcal{F}_\mathrm{p}(\Lambda_\mathrm{full})$ & 67.4\% & 67.7\% & 64.2\% \\ 
    $\mathcal{F}_\mathrm{g}(\Lambda_{-p})$ & 89.5\% & 88.5\% & 87.1\% \\ 
    $\mathcal{F}_\mathrm{g}(\Lambda_{-m})$ & 89.4\% & 87.2\% & 86.0\% \\
    \hline \hline
    \end{tabular}
    \\\vspace{0.2cm}
    \begin{tabular}{cccc} 
    \multicolumn{4}{c}{Identity}
    \\
    \hline \hline
    Fidelity & Standard & Least-square & ML \\ 
    \hline 
    $\mathcal{F}_\mathrm{p}(\Lambda_\mathrm{SPAM})$ & 68.6\% & 71.1\% & 69.8\% \\ 
    \hline \hline
    \end{tabular}
\end{table}

\subsection{Handling SPAM errors}

Through similar considerations we can calculate the net channel of SPAM errors by performing tomography on the identity gate, acquiring the outcome matrix
\begin{eqnarray}
    P^I_{i,j} = \sbra{\rho_i} \Lambda_m \Lambda_p \sket{\rho_j}. 
    \label{eq:PI_SPAM2}
\end{eqnarray}
The physical channel obtained from this process $\Lambda_\mathrm{SPAM} = \Lambda_m \Lambda_p$ includes both state preparation and measurement errors. Since the separation between measurement and state preparation errors is not unique, we consider two limiting cases: $\Lambda_\mathrm{SPAM} = \Lambda_m$ and $\Lambda_\mathrm{SPAM} = \Lambda_p$.

Note, that we assumed that SPAM errors are caused by a single physical channel ($\Lambda_\mathrm{SPAM} \in \mathcal O_\mathrm{CPTP}$), independent from the prepared and measured states. As we pointed out earlier, this is not generally the case and Ref.\ \cite{blume2024easy} proposes a SPAM correction method allowing for more general input and projector states. However, explicitly enforcing physical constraints in this formalism is generally not possible. Therefore, we assume that the actual states can be written in the form $\Lambda_p\sket{\rho_i}$ ($\Lambda_m^\dagger\sket{\rho_i}$) independent of $i$.

The optimization problem in the case of pure measurement errors is
\begin{equation}
    \Lambda_{-m} = \mathrm{argmin}_{\Lambda} C(P^G,A^\dagger \Lambda_m \Lambda A)\, |\,
    \Lambda \in \mathcal O_\mathrm{CPTP},
\end{equation}
where the cost function and how the CPTP constraint is enforced depend on the optimization method described above. Analogously, one can obtain a SPAM-corrected channel $\Lambda_{-p}$, where only state-preparation contributes to SPAM.

The gate fidelities obtained from the SPAM-corrected channels are shown in \tabref{tab:fidelity_compare} for both types of iSWAP gates. The gate fidelities obtained from the SPAM-corrected channels are in better agreement with the RB fidelity, and only marginally differ between different optimization methods. Importantly, not only the fidelities but also the SPAM-corrected channels themselves are quite similar. The process fidelity for any pair of ($\Lambda_{-p}$,$\Lambda_{-m}$) channels is above $0.95$.

\subsection{The unitary component of the channel}

Having the SPAM-corrected quantum channels we can extract the unitary component that reveals the calibration errors of the iSWAP gate. To this we abandon the super-Dirac notation and rewrite the channel in the chi-matrix representation. We define how the channel acts on the density matrix $\rho$ as follows
\begin{eqnarray}
    \Lambda(\rho) &= \frac 1 4 \sum \limits_{m,n} \chi_{mn} P_m\rho P_n,
\end{eqnarray}
where $P_m$ is the $m$th two-qubit Pauli-matrix, and $\chi_{mn}$ is a real symmetric matrix. Note that Pauli matrices here are used in a different context than for the PTM representation and the transformation between the two representations is not trivial. Moving to the eigenbasis of the $\chi_{mn}$ matrix
\begin{eqnarray}
    \Lambda(\rho) &= \frac 1 4  \sum \limits_{i} \lambda_i K^{}_i\rho K^\dagger_i,
\end{eqnarray}
where $K_i = \sum \limits_m v_{im}P_m$ are the Kraus operators of equal norm with $v_{mi}$ being the $m$th component of the $i$th eigenvector.

\begin{table}
    \centering
    \caption{Gate fidelity $\mathcal{F}_\mathrm{g}$ of the unitary component of the iSWAP gate using different assumptions on SPAM and different optimization methods. }
    \label{tab:Unitary_fidelity_compare}
    \begin{tabular}{cccc} 
    \multicolumn{4}{c}{iSWAP (adiabatic), unitary component}
    \\
    \hline \hline
    Fidelity & Standard & Least-Square & ML \\ 
    \hline 
    $\mathcal{F}_\mathrm{g}(U_{-p})$ & 95.5\% & 96.5\% & 96.8\% \\
    $\mathcal{F}_\mathrm{g}(U_{-m})$ & 94.4\% & 95.4\% & 95.8\%\\ 
    \hline \hline
    \end{tabular}
    \\\vspace{0.2cm}
    \begin{tabular}{cccc} 
    \multicolumn{4}{c}{iSWAP (diabatic), unitary component}
    \\
    \hline \hline
    SPAM & Standard & Least-Square & ML \\ 
    \hline 
    $\mathcal{F}_\mathrm{g}(U_{-p})$ & 97.4\% & 97.6\% & 97.9\%\\ 
    $\mathcal{F}_\mathrm{g}(U_{-m})$ & 97.0\% & 97.0\% & 97.2\% \\
    \hline \hline
    \end{tabular}
\end{table}

Since our channel is composed of unitary and stochastic components (deviating from identity with some small error probability), the Kraus operator corresponding to the largest eigenvalue $\lambda_0$ should correspond to the unitary part of the channel. In reality, we have to impose unitarity by
\begin{eqnarray}
    U = (K^{}_0 K_0^\dagger)^{-1/2}K^{}_0,
\end{eqnarray}
where $(K^{}_0 K_0^\dagger)^{-1/2}$ is sufficiently close to the identity \cite{keller1975closest}. E.g., for $\Lambda_{-m}$ of the diabatic gate calculated with the ML method, $\mathcal F_\mathrm{p}(U,K_0) = 0.996$. The full channel can then be approximated as
\begin{eqnarray}
    \Lambda(\rho) \approx \frac{\lambda_0} 4  U\rho U^\dagger + \frac 1 4\sum \limits_{i = 1} \lambda_i K^{}_i\rho K^\dagger_i,
\end{eqnarray}
where $p_\mathrm{incoh} = 1- \lambda_0/4$ can be interpreted as the incoherent error probability. From the calculated gate fidelities of the unitary components in \tabref{tab:Unitary_fidelity_compare} we see that the calibration errors, i.e., $1-\mathcal F_\mathrm{g}(U)$, are of the order of single-qubit gate errors.

\begin{figure*}[tbp]
    \centering
    \includegraphics[width=0.99\textwidth]{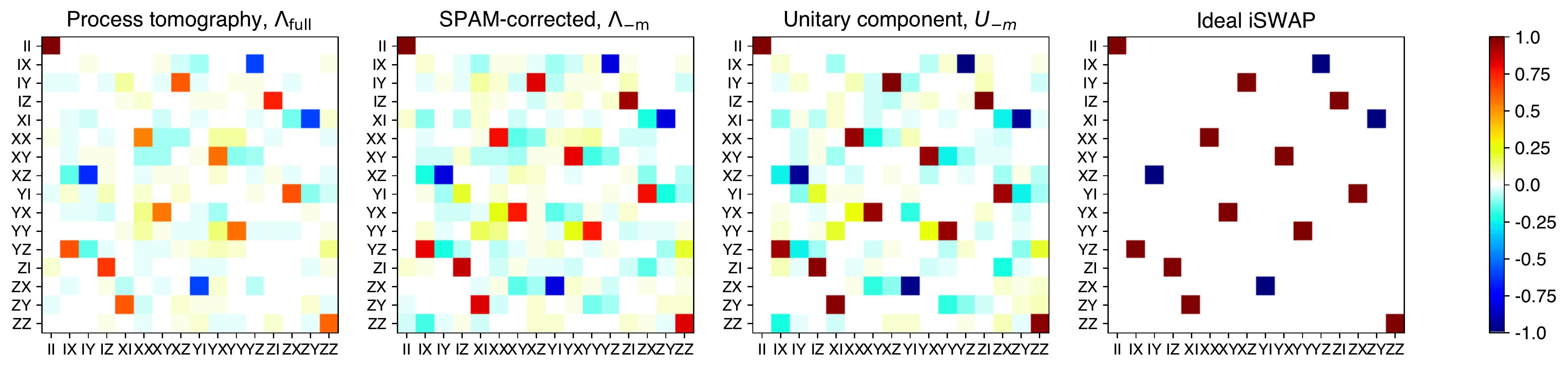}
    \caption{Pauli transfer matrices of the full (physical) channel obtained from standard process tomography; the channel where SPAM errors are assumed to act during the measurement process; the unitary component of the SPAM-less channel; and the channel of the ideal iSWAP gate. The process fidelity between the different transformations are $\mathcal F_\mathrm{p}(\Lambda_\mathrm{full},\Lambda_{-m}) = 0.82$ and $\mathcal F_\mathrm{p}(\Lambda_{-m}, U_{-m}) = 0.85$ with the latter being approximately the $1-p_\mathrm{incoh}$.}
    \label{fig:fitted_channel_mle}
\end{figure*}

Finally, we present the complete fitting procedure on \figref{fig:fitted_channel_mle}. Comparing the different channels we see that removing SPAM errors account for $17.5\%$ of the errors in the process tomography, while incoherent errors ($p_\mathrm{incoh}\approx 14.4\%$) dominate the gate infidelity of the SPAM-corrected channel.

\subsection{Tomography calibration}
\label{app:subsec2}

\begin{figure*}[tbp]
\includegraphics[width=0.75\textwidth]{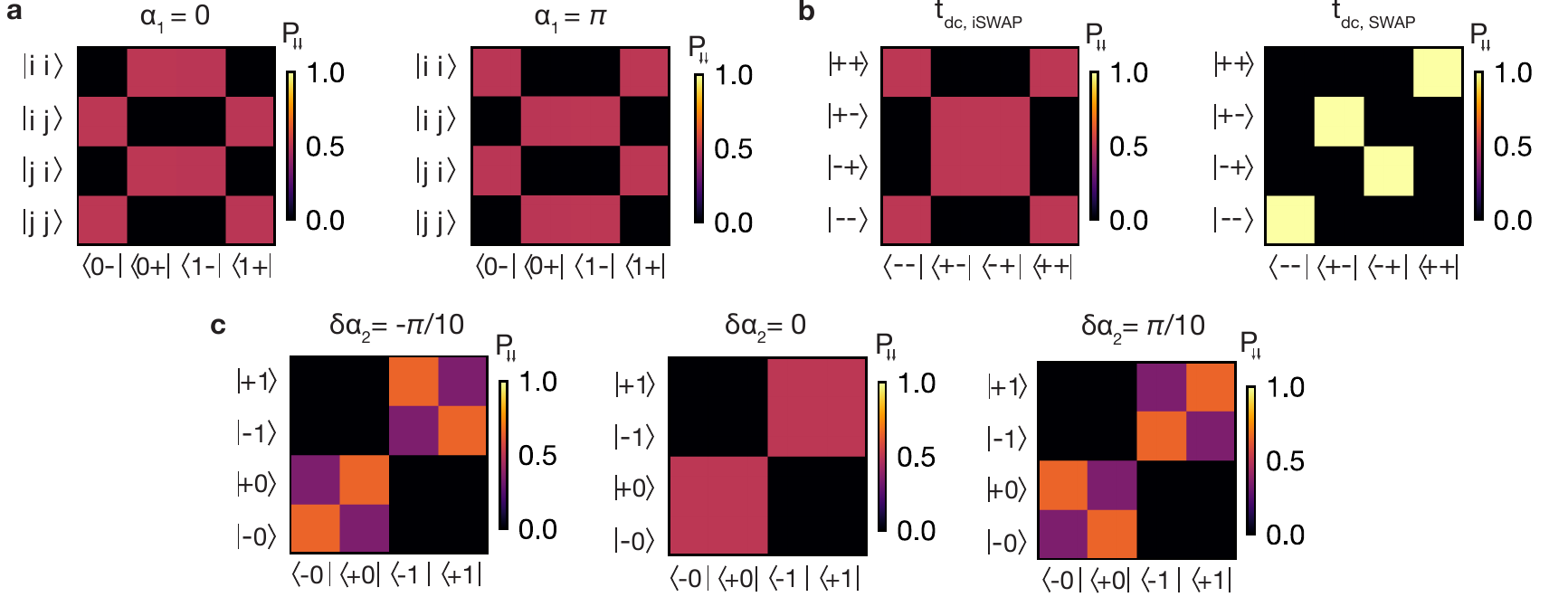}
\caption{Simulated examples of partial tomographies used for iSWAP gate phase tuning. (a) Tomography used to double-check if the phase $\alpha_1$ is wrong by $\pi$. (b) Tomography used to calibrate $t_{\mathrm{dc}}$. If the baseband pulse duration is wrong then the pattern becomes more similar to the SWAP. (c) Tomography used for the calibration of the phase $\alpha_2$. A slight error in $\theta_2$ results in a checkerboard pattern in the off-diagonal blocks, here shown for errors of $\pm\pi/10$. An analogous pattern can be found for $\alpha_1$ if the readout and initialization bases are swapped.} 
\label{fig:partial}
\end{figure*}

After extracting the approximate gate parameters $t_\mathrm{ac}$ and $t_\mathrm{dc}$ and single-qubit phases $\theta_1$ and $\theta_2$ with the Ramsey sequences, these parameters are further fine-tuned using tailored tomographic sequences. Firstly, $t_\mathrm{dc}$ is tuned by preparing the states $\{\ket{--},  \ket{-+}, \ket{+-}, \ket{++}\}$, performing the gate ($U_{\mathrm{fSim}}$), and projecting back onto the same states $\{\ket{--},  \ket{-+}, \ket{+-}, \ket{++}\}$. The resulting pattern changes as $t_\mathrm{dc}$ is tuned between a SWAP and and iSWAP gate, as depicted in the calculation in \figref{fig:partial}b. Importantly, this calibration is insensitive to errors in $\theta_1$ and $\theta_2$, which only affect the contrast of the pattern. For tuning $\alpha_1$ we prepare in the $\{\ket{0-}, \ket{0+}, \ket{1-}, \ket{1+} \}$ basis states and project onto the states: $\{\ket{-0}, \ket{+0}, \ket{-1}, \ket{+1} \}$. We sweep the value of $\alpha_1$ and visually determine the angle that results in the best correspondence with theory. Small phase deviations result in a characteristic checkerboard pattern (\figref{fig:partial}c). Similarly, to determine $\alpha_2$ we prepare in $\{\ket{-0}, \ket{+0}, \ket{-1}, \ket{+1} \}$ and measure $\{\ket{0-}, \ket{0+}, \ket{1-}, \ket{1+} \}$. We note that this process may be inconvenient for distinguishing between $\alpha_{i}$ and $\alpha_{i} + \pi$, since the calibrated result looks the same. This ambiguity can be resolved for $\alpha_1$ by performing a partial tomography where the qubits are initialized in the basis states $\{\ket{jj},  \ket{ji}, \ket{ij}, \ket{ii}\}$ and reading out in the states $\{\ket{-0}, \ket{+0}, \ket{-1}, \ket{+1} \}$ (\figref{fig:partial}a) or, similarly for $\alpha_2$ by preparing $\{\ket{jj},  \ket{ji}, \ket{ij}, \ket{ii}\}$ and read out in the states $\{\ket{0-}, \ket{0+}, \ket{1-}, \ket{1+} \}$. Finally, we note that a conceptually similar tomographic calibration approach, involving sweeps of the single qubit Z corrections as done in this work, could be used to determine the parameters of an arbitrary fSim($\gamma, \zeta$) gate \cite{arute2020observationseparateddynamicscharge}.

\begin{table}
    \centering
    \caption{Final gate parameters}
    \label{tab:gateParams}
    \begin{tabular}{|c|c|c|}
        \hline\hline
        Parameter & Diabatic & Adiabatic \\
         \hline 
       $\mr{vB}_{\text{dc}}$& -14 mV & -12.3 mV \\
       $\mr{vB}_{\text{ac}}$& 4.3 mV & 5.3 mV \\
       $f_{\text{SWAP}}$& 15.9 MHz & 16.3 MHz \\
       $\Delta_{1}$& 6.23 MHz& - \\
       $\Delta_{2}$&  -1.69 MHz & - \\
       $\theta_{1}\mod{2\pi}$&- & 0.90 rad \\
       $\theta_{2}\mod{2\pi}$&- &  2.68 rad  \\
       $t_{\text{dc}}$  &  333 ns & 285 ns\\
       $t_{\text{ac}}$  &  290 ns & 253 ns\\
       $t_{\mathrm{ramp}}$& 5 ns & 100 ns \\
    \hline\hline
    \end{tabular}
\end{table}

\section{Randomized benchmarking}
\label{app:IRB}

\subsection{Single-qubit RB}

Single-qubit randomized benchmarking (RB) is done using the Clifford gates in Table \ref{tab:Single_RB} \cite{epstein2014investigating,hendrickx2024sweet}. The average number of gates per Clifford is 2.125, and the percentage of physical gates is 0.392 (see Table\,\ref{tab:Single_RB_counts}). The identity gate is kept for accounting purposes and is defined as doing nothing for no time. Average physical gate infidelity is calculated as $(1-F_{\text{C}})/2.125 / 0.392$, where $F_{\text{C}}$ is the average Clifford gate fidelity. The provided $1\sigma$ uncertainty is computed with respect to the nonlinear least squares fit as the square root of the corresponding diagonal element of a covariance matrix (using scipy.curvefit).

\begin{table}[tbp]
    \centering
    \caption{Implementations of all single-qubit Clifford gates \cite{hendrickx2024sweet}. There are 2.125 gates per Clifford on average.}
    \label{tab:Single_RB}
\begin{tabular}{|>{\bfseries}c|}
\hline
Clifford gates \\
\hline
$\mathrm{I}$ \\
\hline
$\mathrm{-Z},\, \mathrm{X}$ \\
\hline
$\mathrm{-Z},\, \mathrm{X}/2$ \\
\hline
$\mathrm{-Z},\, \mathrm{X}/2,\, \mathrm{-Z}$ \\
\hline
$\mathrm{-Z},\, \mathrm{X}/2,\, \mathrm{-Z}/2$ \\
\hline
$\mathrm{-Z},\, \mathrm{X}/2,\, \mathrm{Z}/2$ \\
\hline
$\mathrm{-Z}/2$ \\
\hline
$\mathrm{-Z}/2,\, \mathrm{X}/2$ \\
\hline
$\mathrm{-Z}/2,\, \mathrm{X}/2,\, \mathrm{-Z}$ \\
\hline
$\mathrm{-Z}/2,\, \mathrm{X}/2,\, \mathrm{-Z}/2$ \\
\hline
$\mathrm{-Z}/2,\, \mathrm{X}/2,\, \mathrm{Z}/2$ \\
\hline
$\mathrm{X}$ \\
\hline
$\mathrm{X},\, \mathrm{Z}/2$ \\
\hline
$\mathrm{X}/2$ \\
\hline
$\mathrm{X}/2,\, \mathrm{-Z}$ \\
\hline
$\mathrm{X}/2,\, \mathrm{-Z}/2$ \\
\hline
$\mathrm{X}/2,\, \mathrm{Z}/2$ \\
\hline
$\mathrm{Z}$ \\
\hline
$\mathrm{Z}/2$ \\
\hline
$\mathrm{Z}/2,\, \mathrm{X}$ \\
\hline
$\mathrm{Z}/2,\, \mathrm{X}/2$ \\
\hline
$\mathrm{Z}/2,\, \mathrm{X}/2,\, \mathrm{-Z}$ \\
\hline
$\mathrm{Z}/2,\, \mathrm{X}/2,\, \mathrm{-Z}/2$ \\
\hline
$\mathrm{Z}/2,\, \mathrm{X}/2,\, \mathrm{Z}/2$ \\
\hline
\end{tabular}
\end{table}

\begin{table}[]
    \centering
    \caption{Gates composing the Clifford group. Fraction of physical gates is $0.392$.}
    \label{tab:Single_RB_counts}
    \begin{tabular}{|c|c|c|}
    \hline
    \textbf{Gate Type}  & \textbf{Percentage} \\
    \hline
    Z/2, Z & 58.82\% \\
    X/2 & 31.37\% \\
    X & 7.84\% \\
    I & 1.96\% \\
    \hline
    \end{tabular}
\end{table}

\subsection{Interleaved RB}

For interleaved randomized benchmarking (IRB) we prepare in $|{\downarrow\downarrow}\rangle$ and apply a random sequence of two-qubit Clifford gates composed of single-qubit gates. The Clifford group is generated from the set shown in Table\,\ref{tab:Single_RB}. From this it follows that the recovery gate of the reference sequence is always composed of single-qubit gates. However, the recovery gate of the interleaved sequence may include multiple iSWAP gates, leading to a large horizontal offset between the decay curves. In order to minimize the number of iSWAP gates in the recovery gate, we transpile it using the gate set $\{Z(\theta), \text{X}, \text{X}/2, \text{iSWAP}\}$, which now includes fractional virtual $Z(\theta)$ rotations. The transpilation is done using the Qiskit "transpile" function.

\section{Experimental setup}
\label{app:setup}

The measurements are performed in a Bluefors XLD dilution refrigerator with a base temperature of approximately 20 mK. All RF signals are generated using Tektronix
AWG5024C arbitrary waveform generators. All RF signals used for the gate are generated on the same (master) AWG. A channel of a second (slave) AWG is used for the virtualization of the SHT plunger. The sensor conductance is measured using two Basel Precision Instruments SP983c IV-converters with a gain of $10^6$ V/A and a low-pass output filter with a cut-off frequency of 30 kHz. The source-drain bias is set to
250 µV. The outputs of the IV converters are fed to a BasPI SP1004 differential amplifier with a gain of $10^2$ and the signal is read using an Alazar ATS9440
digitizer card. The digitizer card can either return a result averaged over shots, or the full sequence of shot outcomes. 

The calibration of the probability scale is done using one of the following methods:
\begin{enumerate}
    \item \textbf{Averaged time-resolved current measurement with I/X$_\pi^{\mathrm{Q}1}$ normalization} (\figref{fig1}, \ref{fig2}, \ref{fig3}b, \ref{fig3}c. and \ref{fig:fig2}). 
    Shots are integrated directly on the digitizer and an approximate probability scale is determined from measuring the averaged outcomes of $|{\downarrow\downarrow}\rangle$  and $|{\uparrow\downarrow}\rangle$ states prepared in a dedicated pre-pended or post-pended experiment. This procedure removes SPAM errors.  
    \item \textbf{Averaged time-resolved current measurement with probability scale normalized based on a neighboring single-shot measurement} (\figref{fig:fig4}a). Same as above, but the normalization is now done by fitting a double-gaussian to the histogram of shots in a dedicated pre-pended or post-pended experiment. SPAM errors are therefore not removed.
    \item \textbf{Single-shot measurement } (\figref{fig3}d, \ref{fig:fig4}b and \ref{fig:fitted_channel_mle}). Here all shots are collected. A threshold is determined by taking the average for outcomes of $|{\downarrow\downarrow}\rangle$  and $|{\uparrow\downarrow}\rangle$. This form of readout is used for all of the partial tomographies and full quantum process tomographies. 
\end{enumerate}

\end{document}